\documentclass{aa}  

\usepackage{graphicx}
\usepackage{subfig}
\usepackage{txfonts}
\usepackage{float}
\usepackage{amssymb}
\newcommand{\grad}{$^{\circ}$}

\begin{document}

   \title{A model for microquasars of Population III}

   \author{Sotomayor Checa, P.\thanks{Fellow of CONICET.}\fnmsep
          \inst{1,2}
          \and
          Romero, G.E.\thanks{Member of CONICET.} \inst{1,2}
          }

   \institute{Instituto Argentino de Radioastronom\'{\i}a (CCT-La Plata, CONICET; CICPBA), C.C. No. 5, 1894,Villa Elisa, Argentina\\
              \email{psotomayor@iar.unlp.edu.ar}
         \and
            Facultad de Ciencias Astron\'omicas y Geof\'{\i}sicas, Universidad Nacional de La Plata, Paseo del Bosque s/n, B1900FWA La Plata, Argentina\\
             }

   \date{Received ***; accepted ***}

 
  \abstract
   {Current simulations indicate that the first stars were formed predominantly in binary systems. The study of the contribution of the first accreting binary systems to the reionization and heating of the intergalactic medium requires the formulation of a concrete model for microquasars of Population III.}
   {We aim at constructing a complete model for microquasars where the donor star is of Population III.}
   {We consider that the  mas-loss of the star is caused exclusively by the spill of matter through the Roche lobe towards the black hole. We calculate the spectral energy distribution of the radiation produced by the accretion disk, the radiation-pressure driven wind, and the relativistic particles in the jets, within the framework of a lepto-hadronic model. In addition, we estimate the impact on the reionization by the first microquasars.}
   {We determine that Population III microquasars are powerful sources of ultraviolet radiation produced by the winds of their super-critical disks, and generate a broadband non-thermal emission in jets.}
   {Our results indicate that microquasars in the early Universe could have been important for the reionization and heating of the intergalactic medium.}
    

   \maketitle
%

\section{Introduction}
After the epoch of recombination, when the first neutral atoms formed, the universe underwent a period of darkness known as the "dark ages" \citep{ellis2012}. During this period there were only two sources of light: the cosmic microwave background (CMB) produced in the decoupling, and the hyperfine transitions of the neutral hydrogen atoms. The radiating structures did not exist yet. When the first stars formed, the intense ultraviolet radiation emitted by them ionized the neutral intergalactic medium (IGM) \citep[see][]{loeb2010}, giving rise to the period known as reionization. This phase transition of matter marked the end of the dark ages. However, current research indicates that the radiation of the first stars alone would not have been sufficient to ionize the intergalactic medium on long distances because of the high columnar density of the clouds in which they formed. Accreting sources powered by intermediate-mass and supermassive black holes have been proposed to solve this problem \citep[see][and~references~therein]{madau2004,bosch-ramon2018}, because the mean free path of the X-rays is longer than that of the predicted ultraviolet photons of the stars of Population III \citep[see][]{bromm2013}. Recent studies indicate, however, that these sources would not have contributed significantly to the reionization process \citep[see][]{milosavljevic2009,qin2017}. It becomes necessary, then, to explore other alternatives for the reionization.\par
At the beginning of this decade, \cite{mirabel2011} suggested that the first high-mass X-ray binaries might have played an important role in the termination of the dark ages. The existence of these sources at high redshift is in agreement with current studies indicating that the first stars were born predominantly in binary systems \citep[see][and~references~therein]{bromm2013}. 
Microquasars are X-ray binaries with relativistic jets \citep{mirabel1999}. These sources can generate gamma-rays in the interaction between accelerated particles in the jets and magnetic, matter or radiation fields \citep[see][]{bosch-ramon2006,reynoso2008b,romero2008,romero2014}. \cite{tueros2014} studied the contribution of cosmic rays accelerated in the terminal regions of the jets in the first microquasars to the reionization of the IGM. Microquasars as alternative sources for the heating of the IGM during reionization of the universe have also been studied recently by \cite{douna2018}. These suggestions, although attractive and energetically consistent, are not supported by a specific model of microquasars of Population III. These microquasars must have characteristics different from those known in the Galaxy which are of Population I and II, since donor stars present a distinct physics caused by the lack of metallicity \citep{bahena2010,bahena2012}.\par 
The primary goal of this paper is to present a complete model for microquasars of Population III, in order to provide a tool for quantitative predictions. As a secondary objective we intend to make realistic estimates of the production of radiation and cosmic rays that will be injected into the intergalactic medium by these objects. We adopt for the binary system the physical parameters provided by  \cite{inayoshi2017} and numerically calculate some relevant physical properties of the accretion disk. A powerful wind is ejected from the accretion disk, driven by radiation pressure. In the innermost region of the disk we assume that the jet launching takes place by a magneto-centrifugal mechanism. In a region close to the compact object, charged particles can be accelerated up to relativistic energies by internal shocks \citep[see][]{drury1983,spada2001}. We adopt a lepto-hadronic model to calculate the radiation produced in the jets by the interaction of the relativistic particles with the different ambient fields. In addition, we study the production of radiation by the cooling of accelerated electrons in the terminal region of the jet \citep[as~considered,~e.g.,~by][]{bordas2009}. All calculations have been computed for a cosmological redshift value $z=10$. Preliminary results have been presented in \cite{romero2018} and \cite{sotomayor2019a}.\par
The paper is organized as follows. In Sect. \ref{sect:model} the model is presented. We consider for the accretion disk parameters with physically acceptable values in order to make quantitative predictions. In the study of the jets, we propose four regions of particle acceleration: a region close to the compact object, and three emission zones due to the interaction of the jets with their environment (bow-shock, cocoon and reconfinement region). In Sect. \ref{sect:results} we present the results of our calculations. We include the effects of absorption by external and internal photon-photon annihilation on the spectrum and we show the total radiative output of the system. In Sect. \ref{sect: reionization} elementary notions on ionization of IGM by winds and jets of microquasars of Population III are discussed. Section \ref{sect:discussion} contains a discussion of the results. Section \ref{sect:conclusions} is devoted to concluding remarks and the prospects of our model for future applications.
\section{Model}
\label{sect:model}
\subsection{General considerations}
\label{subsect:general_considerations}
We consider a binary system whose components are a Population III star of $M_{*}=41\,M_{\odot}$ and a black hole of $M_\mathrm{BH} = 34\,M_{\odot}$. The orbital separation of the components is $a=36\,R_{\odot}$\footnote{The masses and the orbital separation correspond to the binary system of Population III consistent with the gravitational wave source GW150914 detected by LIGO \citep{inayoshi2017}.}. The star radius is approximated by the average radius of its Roche lobe  \citep{eggleton1983}:
\begin{equation}
R_{*}\approx R_{\mathrm{L}}\approx a \frac{0.49q^{2/3}}{0.6q^{2/3}+\ln \left(1+q^{1/3}\right)},
\label{RadioRoche}
\end{equation}
where $q=M_{*}/M_{\mathrm{BH}}$. We assume the orbit to be circular, which is a good approximation in binary systems with mass transfer \citep[see e.g.][]{paczynski1971}.\par
Unlike what happens in massive stars in the Galaxy, the mass-loss by stellar winds is an inefficient mechanism for Population III stars \citep[see][and references therein]{krtica2006}. Instead, the mass-transfer is exclusively by overflowing of the Roche lobe. We assume that the black hole accretes matter of the donor star by spilling through the Lagrange point $\mathrm{L_{1}}$. \par
\begin{figure}[h!]
  \centering
  \includegraphics[scale=0.48]{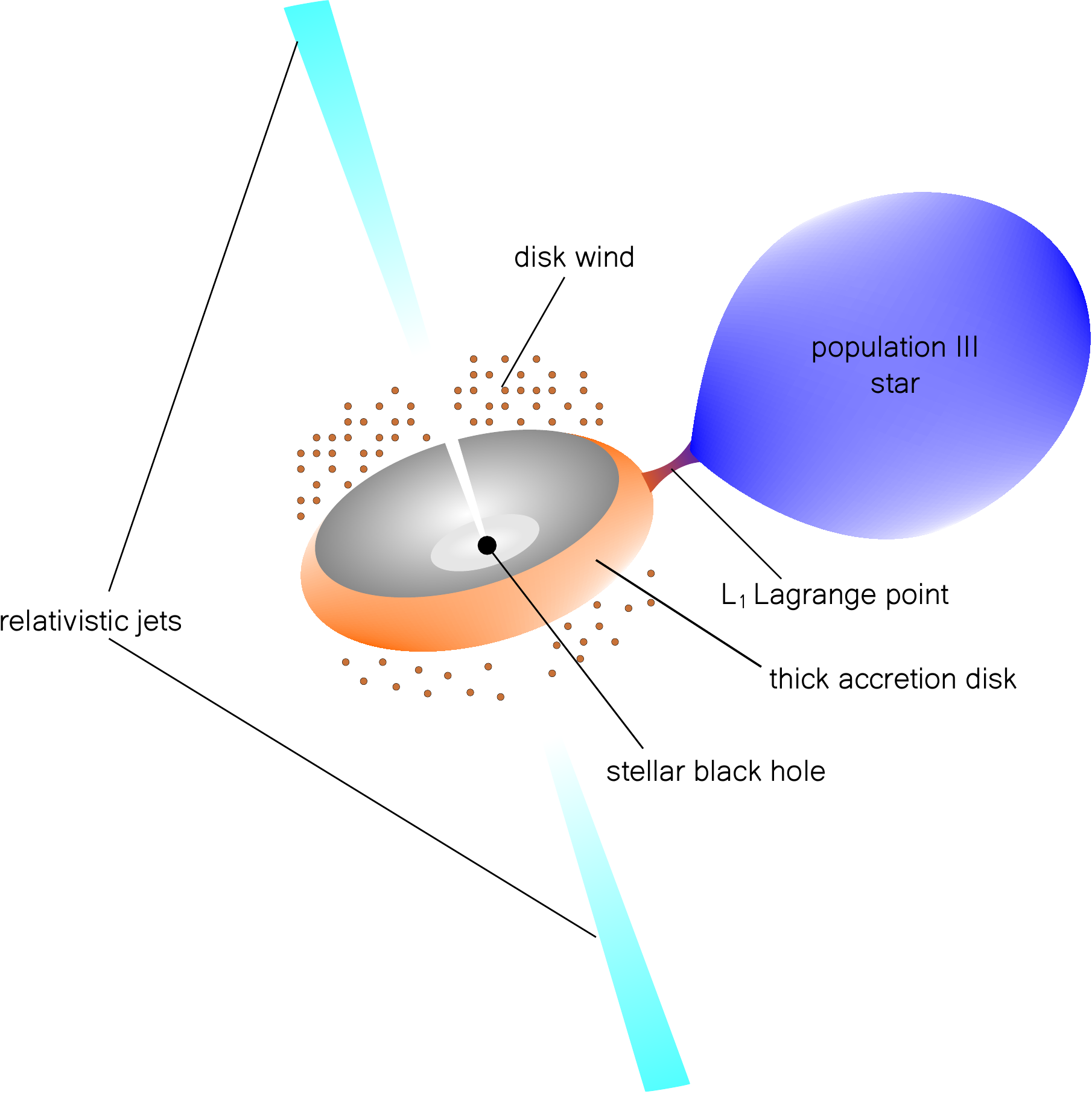}
     \caption{\small Scheme of the Population III microquasar. When the star fills its Roche lobe the black hole accretes matter forming an accretion disk around the compact object. The rate of accretion is very intense and the disk becomes thick, expelling an intense wind of particles from the disk. From the innermost region of the disk, powerful relativistic jets are ejected. }
        \label{fig:microquasar}
\end{figure}
The black hole accretes matter until the donor star becomes a black hole by direct collapse \citep[see][]{heger2003}, giving rise to a binary system of two black holes. Calculations by \citet{inayoshi2017} show that the life-time of such a system in the microquasar phase is $\tau_{\mathrm{MQ}}\sim 2\times 10^{5}\,\mathrm{years}$. The masses of the components before the second collapse are $M_{*} \approx 26\,M_{\odot}$ and $M_{\mathrm{BH}} \approx 36\,M_{\odot}$. Then, the star loses $15\,M_{\odot}$ but the mass of the black hole increases by $2\,M_{\odot}$. The latter indicates that most of the accreted matter is expelled from the system. We consider the mass-transfer is stable and has a constant rate given by:
\begin{equation}
\dot{M}_{*}\sim \frac{M_{*,i}-M_{*,f}}{\tau_{\mathrm{MQ}}}.
\label{eq:loss-mass}
\end{equation}
From Eq. (\ref{eq:loss-mass}) we find that the accretion rate at the outer edge of the accretion disk is much higher than the critical accretion rate of the black hole ($\dot{M} \gg \dot{M}_{\mathrm{crit}} = 4\pi GM_{\mathrm{BH}}m_{\mathrm{p}}/\sigma_{\mathrm{T}}c$), therefore the accretion regime is super-critical (super-Eddington). To model the accretion disk we cannot adopt the classical standard case developed in \cite{shakura1973}. Instead, we consider a thick accretion disk similar to the one observed in SS433, the only microquasar known in the Galaxy in a super-Eddington accretion regime \citep{calvani1981}. \par
Powerful, highly collimated, oppositely directed relativistic jets are ejected from the innermost region of the accretion disk \citep{mirabel1994}. If the jet is composed of accreted matter, it must contain both electrons and protons, as well as a magnetic field associated with the plasma. In this paper we consider hadronic jets, as observed in SS433 \citep{fabrika2004}. For simplicity, we assume that the jets are perpendicular to the orbital plane and that they do not precess. In addition, we do not distinguish between the jet and the counterjet.\par
The jet begins to decelerate when the ram pressure of the environment is balanced with the momentum flux carried by the jet head. Then, two shocks form in the head of the jet: a bow-shock propagating in the external medium and a reverse-shock directed towards the interior of the jet. The matter that crosses the reverse shock inflates a region of the jet known as the cocoon. At the point of the jet where the pressure is equal to the pressure of the cocoon, another shock called recollimation shock is formed. For the calculation of radiative processes in jets we consider acceleration of particles in a region close to the compact object (inner jet), and in the terminal region of interaction with the environment (terminal jet). The jet model and the radiative processes considered here are based on the models developed by \cite{bosch-ramon2006,reynoso2008b,romero2008,bordas2009}; and \cite{vilatesis}.\par
A sketch of the system is shown in Fig. \ref{fig:microquasar}. In Table \ref{table: OrbitalParameters} we show the main parameters of the binary adopted in this paper.
\begin{table}[h]
\caption{\small Parameters of the binary system}             
\label{table: OrbitalParameters}      
\centering                          
\begin{tabular}{c c c c}        
\hline\hline                 
Parameter & Symbol & Value \\    
\hline                        
   Donor star mass \small $^{\dagger}$& $M_{*}$ & $41$ $M_{\odot}$ \\     
   Black hole mass \small $^{\dagger}$& $M_{\mathrm{BH}}$ & $34$ $M_{\odot}$ \\
   Star radius \small $^{\ddagger}$& $R_{*}$ & $14.2$ $R_{\odot}$ \\
   Star luminosity \small $^{\dagger}$& $L_{*}$ & $1\times 10^{6}$ $L_{\odot}$ \\
   Star temperature \small $^{\dagger}$& $T_{*}$ & $5\times 10^{4}$ K \\ 
   Orbital semiaxis \small $^{\dagger}$& $a$ & $36$ $R_{\odot}$ \\
   Orbital period \small $^{\ddagger}$& $P$ & $2.9$ days \\
   Mass loss rate \small $^{\ddagger}$& $\dot{M}_{\mathrm{*}}$ & $7.5\times 10^{-5}$ $M_{\odot}\,\mathrm{yr}^{-1}$ \\
   Eddington rate \small $^{\dagger}$& $\dot{M}_{\mathrm{Edd}}$ & $2.2\times 10^{-8}$ $M_{\odot}\,\mathrm{yr}^{-1}$ \\

   Eddington luminosity \small $^{\dagger}$& $L_{\mathrm{Edd}}$ & $4.3\times 10^{39}$ $\mathrm{erg}\,\mathrm{s}^{-1}$ \\
   Gravitational radius \small $^{\dagger}$& $r_{\mathrm{g}}$ & $50$ $\mathrm{km}$ \\
\hline                                   
\
\small $^{\dagger}$ fixed \\
\small $^{\ddagger}$ calculated
\end{tabular}
\end{table}
\subsection{Disk and wind}
\label{subsect:accretion_disk}
%
At the super-Eddington regime, the accretion disk must become thick, at least in the region close to he compact object \citep[see][]{jaroszynski1980,paczynski1980}. Considering the prescription developed by \cite{shakura1973} we can characterize the efficiency of the angular transport mechanism by just one parameter $\alpha$. \cite{jaroszynski1980} showed that in a thick accretion disk in hydrostatic equilibrium the viscosity parameter should be such that $\alpha \ll 1$. We adopt the value $\alpha = 0.01$, constant along the disk in our calculations.\par
The main characteristic of super-critical accretion disks is the photon-trapping \citep[see e.g.][]{begelman1978,ohsuga2003,ohsuga2005}. Photon-trapping occurs when the photon diffusion time-scale exceeds the accretion time-scale. Under such circumstances, photons generated via the viscous process are advected inward with the accreted flow without being able to go out from the surface. Following \cite{narayan1994}, the photon-trapping effect can be parameterized considering that advection heating is a fraction of the viscous heating, $Q_{\mathrm{adv}} = f Q_{\mathrm{vis}}$, where $f$ is the advection parameter, constant along the disk.\par
Toroidal magnetic fields are expected to develop in the accretion disk of X-ray binaries, where the stream overflowing the Roche lobe would stretch out in the toroidal and radial directions as it feeds the outer disk, and the disk shear would then substantially amplify the toroidal component \citep{liska2018}. However, it is expected that on long distances from the compact object, the dynamics of the plasma is determined mainly by gravity and radiation pressure. Winds are launched within the region where the radiative force overcomes gravity, which occurs inside the critical radius given by \citep[see Eq. (3) of][]{fukue2004}:
\begin{equation}
r_{\mathrm{cr}}=\frac{9\sqrt{3}\sigma_{\mathrm{T}}}{16\pi m_{\mathrm{p}}c}\dot{M}_{\mathrm{*}}.
\label{eq:rcrit}
\end{equation}
We consider that the wind is ejected between the critical radius $r_{\rm cr} \approx 3600\,r_{\rm g}$, and an inner radius given by $r_0 = 100\,r_{\rm g}$. Inside the inner radius, the jets are launched (see Sect. \ref{subsect:jets}).
Similarly to the treatment developed by \cite{narayan1994} and \cite{akizuki2006}, we assume a steady-state and axisymmetric disk. We solve the equations of the dynamics of the accreted fluid using the self-similar treatment developed by \cite{narayan1994}. We adopt for the exponent of the self-similar solutions a value of $s = 1/2$, and we consider for the advection parameter $f$ two values $f = 0.1$ and $0.5$, similar to those in \cite{fukue2004}. For calculations of the most relevant physical properties of the accretion disk, (thickness, temperature, magnetic field, radial velocity and spectral energy distribution of the radiation emitted) the reader is referred to \cite{watarai1999}, \cite{fukue2004}, and \cite{akizuki2006}.\par
Following \cite{fukue2009} and \cite{fukue2010} we assume that the spherically symmetric wind is highly ionized and blows off at a constant speed $\beta = v/c$, at a constant mass-outflow rate $\dot{M}_{\mathrm{wind}}$, and at a constant comoving luminosity $\dot{e} = \dot{E}/L_{\mathrm{Edd}}$. For the wind speed we consider the escape velocity of the gravitational attraction of the central black hole. In the calculations we adopt $\dot{e} = 1$, which is consistent with what was found by \cite{zhou2018}.\par
The optical depth of the wind is roughly estimated as $\tau_{\rm wind} \sim  \sigma_{\rm T} \dot{M}_{*}/4\pi m_{\rm p} v_{\rm wind}r_{\rm cr}$ \citep[see][]{kitabatake2002}, assuming a constant wind velocity $v_{\rm wind}$
. For a wind velocity similar to that of SS433 the optical depth is $\tau_{\rm wind} \approx 27.5$, whereas if we consider the escape velocity at $r_{\rm cr}$ it is $\tau_{\rm wind} \approx 13.8$, and $\tau_{\rm wind} \approx 2$ at $r_0$. Therefore, the wind is optically thick and locally emits blackbody radiation in the comoving reference frame. In this accretion regime, the emission of the wind would hide that of the disk. In the case of an extended disk, the emission of the star can also be covered, similar to that detected in SS433. The comoving temperature is given by:
\begin{equation}
\frac{4\pi r_{\mathrm{g}}^{2}\sigma_{\mathrm{T}}T_{\mathrm{wind,co}}^{4}}{L_{\mathrm{Edd}}} = \dot{e}\frac{r_{\mathrm{g}}^{2}}{R^{2}},
\label{eq:comoving_temperature_wind}
\end{equation}
where $T_{\mathrm{wind,co}}$ is the wind comoving temperature.\par
The apparent photosphere of the wind is the surface where the optical depth, $\tau$, measured from and observer at infinity, becomes equal to 1. The height of the apparent photosphere from the equatorial plane, $z_{\mathrm{ph}}$, is given by \citep{fukue2010}:
\begin{equation}
\tau_{\mathrm{ph}} = - \int	_{\infty} ^{z_{\mathrm{ph}}} \Gamma_{\mathrm{wind}} \left(1 - \beta \cos \theta \right)\kappa_{\mathrm{co}}\rho_{\mathrm{co}}\mathrm{d}z = 1,
\label{eq:height_photosphere_wind}
\end{equation}
where $\Gamma_{\mathrm{wind}}$ is the Lorentz factor of the wind, $\kappa_{\mathrm{co}}$ is the opacity for Thomson electron scattering, and $\rho_{\mathrm{co}}$ is the mass density. Physical quantities with subscript "${\rm co} $" are measured in the comoving frame.
\par
Once the location of the apparent photosphere is determined, the temperature measured by an observer at infinity can be calculated from the comoving temperature by applying the appropriate boost:
\begin{equation}
T_{\mathrm{wind,obs}} = \frac{1}{\Gamma_{\mathrm{wind}}\,\left(1 - \beta \cos \theta\right)}T_{\mathrm{wind,co}},
\label{eq:observed_temperature_wind}
\end{equation}
where $\theta$ is the viewing angle measured from the $z$-axis. Using this observed temperature, we can calculate the spectral energy distribution of the wind measured from an observer at infinity.\par
\begin{table}[h!]
\caption{\small Parameters of the super-critical wind}             
\label{table_OutflowsParameters}      
\centering                          
\begin{tabular}{c c c c}        
\hline\hline                 
Parameter & Value \\    
\hline                        
   Wind mass-loss rate [$\dot{M}_{\mathrm{wind}}$] & $7.3\times 10^{-5}$ $M_{\odot}\,\mathrm{yr}^{-1}$ \\     
   Power of the wind [$L_{\mathrm{wind}}$] & $4.1\times 10^{42}$ $\mathrm{erg\,s^{-1}}$ \\     
   Wind escape velocity at $r_{\mathrm{cr}}$ [$v_{\mathrm{wind}} (r_{\mathrm{cr}})$] & $6\times 10^{3}$ $\mathrm{km\,s^{-1}}$ \\
   Wind escape velocity at $r_{0}$ [$v_{\mathrm{wind}} (r_{0})$] & $4.2\times 10^{4}$ $\mathrm{km\,s^{-1}}$ \\
\hline                                   
\
\small all parameters are calculated
\end{tabular}
\end{table}
In super-critical accretion disks the launch of jets by radiation pressure has been studied extensively \citep[see e.g.][]{piran1982,ohsuga2005,sadowski2015}. In this work, we adopt a mechanism of magneto-centrifugal launching, which is generally adopted for galactic microquasars \citep{meier2005}. Although the model developed so far does not predict the launch of jets because it includes only toroidal magnetic fields in the accretion disk, we will see below that it is possible to launch the jets, assuming magnetorotational instabilities in the disk.\par
In the Table \ref{table_OutflowsParameters} we show the values of the different parameters calculated for the wind.\par
\subsection{Jets}
\label{subsect:jets}
\subsubsection*{Launching}
\cite{liska2018} showed that under certain physical conditions, a large-scale poloidal magnetic field can be generated from a purely toroidal field using global GRMHD simulations. The toroidal field is relatively strong,  $\beta = p_{\mathrm{gas}}/p_{\mathrm{mag}} = 5$,  to ensure that the simulations resolve the MRI due to both the initial toroidal and dynamo-generated poloidal magnetic fields. The poloidal magnetic flux accumulates around the black hole until it becomes dynamically-important, and leads to a magnetically arrested disc \citep[MAD, see][]{narayan2003}. This disk launches relativistic jets that are more powerful than the accretion flow ($ \zeta = L_{\mathrm{jet}}/L_{\mathrm{acc}} > 1$).\par
The result of \cite{liska2018} can be applied to our model if we consider that the toroidal field has spread throughout the whole disk and after MRI a large-scale poloidal magnetic field is generated. For $M = 34\,M_{\odot}$, the large-scale poloidal magnetic field is generated $t = 7\,\mathrm{s}$ after the magnetorotational instabilities began. The disk is arrested at $r=r_{0}$ and becomes to a MAD. The power of the jet can be calculated from the power accreted at $r=r_{0}$. We adopt the same value for the beta factor of the plasma, $\beta = 5$, the distance to the black hole to which the disk is arrested is $r_{0} = 100\,r_{\mathrm{g}}$, and the power of the jet is determined assuming $\zeta = 1.5$.\par
The jet is injected at $z_{0} = 100\, r_{\mathrm{g}}$ above the accretion disk. The jet luminosity is then given by:
\begin{equation}
\zeta = \frac{L_{\mathrm{jet}}}{L_{\mathrm{acc}}(r_{0})}.
\end{equation}
Therefore, the Lorentz factor of the jet is:
\begin{equation}
L_{\mathrm{jet}} = \Gamma_{\mathrm{jet}}\dot{M}_\mathrm{acc}c^{2},
\end{equation}
where $\dot{M}_\mathrm{acc}$ is the accretion rate at $r_{0}$. Typical values are $\Gamma_{\mathrm{jet}} \sim 10$.\par
\subsubsection*{Inner jet}
In a magneto-centrifugal launching mechanism, the jet is ejected by conversion of magnetic energy into kinetic energy. In general, the magnetic field near the compact object has a higher value than equipartition with the matter of the jet. The magnetic field in the jets decreases with distance z to the compact object. We consider that the flow expands adiabatically, and that:
\begin{equation}
B(z)=B(z_{0})\left(\frac{z_{0}}{z}\right),
\label{equation_MagneticField}
\end{equation}
where $z_{0}$ is the launch point of the jet. The value of $B(z_{0})$ is determined by requiring equipartition between the densities of kinetic and magnetic energy at $z_{0}$:
\begin{equation}
\frac{B^{2}(z_{\mathrm{0}})}{8\pi}=\frac{L_{\mathrm{jet}}}{\pi r_{\mathrm{0}}^{2}v_{\mathrm{jet}}},
\label{equation_EquipartitionDensities}
\end{equation}
where $v_{\mathrm{jet}}$ is the bulk velocity of the jets.\par
The energy density of matter, at any distance on the compact object, is:
\begin{equation}
e_{\mathrm{p}}(z)=\frac{\dot{m}_{\mathrm{jet}}}{2\pi z^{2}}v_{\mathrm{jet}},
\label{equation_MatterDensity}
\end{equation}
and the density of cold protons at a distance $z$ from the black hole in the jet is
\begin{equation}
n_{\mathrm{p}}(z) \simeq \frac{\dot{m}_{\mathrm{j}}}{\pi[R_{\mathrm{j}}(z)]^{2}m_{\mathrm{p}}v_{\mathrm{j}}}.
\end{equation}
These cold protons are targets for the relativistic electrons and protons.\par
Internal shocks produce when the expansion makes the flow compressible, and they convert bulk kinetic energy into random kinetic energy of relativistic particles \citep[see][]{spada2001}. The relativistic protons and electrons that are injected locally are called primary particles. The secondary particles are pions, muons and electron-positron pairs produced as a result of the interaction of primary particles with matter and radiation. In the following, we outline how to calculate the broadband non-thermal emission generated by cooling of all these particles in the inner jet. This procedure is also applicable to calculate the radiation emitted in the terminal jet.\par
We assume an injection function that is a power-law in the energy of the primary particles and that is inversely proportional to the distance to the compact object \citep{romero2008}:
\begin{equation}
Q(E,z) = Q_{\mathrm{0}}\frac{E^{-p}}{z} \qquad \left[Q\right] = \mathrm{erg}^{-1}\mathrm{s}^{-1}\mathrm{cm}^{-3},
\label{equation_InyectionParticles}
\end{equation}
where $p$ is the spectral index of particle injection. The normalization constant $Q_{\mathrm{0}}$ is obtained as:
\begin{equation}
L_{(\mathrm{e,p})} = \int _{V} \mathrm{d}^{3}r \int ^{E_{(\mathrm{e,p})} ^{\mathrm{max}}} _{E_{(\mathrm{e,p})}^{\mathrm{min}}}\mathrm{d}E_{(\mathrm{e,p})}E_{(\mathrm{e,p})}Q_{(\mathrm{e,p})}(E_{(\mathrm{e,p})},z),
\label{equation_LuminosityParticles}
\end{equation}
where $V$ is the volume of the acceleration region. The maximum energy $E^{\mathrm{max}}$ that a relativistic particle can attain is obtained by balancing its rate of acceleration and cooling.\par 
The cooling rate is the sum of the radiative cooling rate and the rate of the adiabatic losses. The radiative losses are caused by the interaction of the relativistic particles with the fields in the jet. We consider electron losses by synchrotron radiation, inverse Compton scattering and relativistic Bremsstrahlung, and proton losses by synchrotron radiation, inelastic proton-proton collisions, and photo-hadronic interactions. The adiabatic losses are caused by the work the particles exert on the walls of the jet. Then \citep{bosch-ramon2006}:
\begin{equation}
t_{\mathrm{ad}}^{-1} = \frac{2}{3} \frac{v_{\mathrm{jet}}}{z}.
\label{equation_AdiabaticRate}
\end{equation}
Detailed formulas for calculating the cooling rate for the radiative processes considered here can be found in \cite{blumenthal1970,stecker1968,begelman1990,atoyan2003,berezinskii1990,kelner2006}; and \cite{romero2014}. The acceleration rate for a charged particle in a magnetic field $B$ is \citep[e.g.][]{aharonian2004}:
\begin{equation}
t_{\mathrm{acc}}^{-1} = \frac{\eta e c B}{E},
\label{equation_AccelerationRate}
\end{equation}
where $E$ is the energy of the particle, and $\eta$ is a parameter that characterizes the efficiency of the acceleration mechanism. The size of the acceleration region is a constraint on the maximum energy if the particle gyroradius $r_{\mathrm{g}}(z)= E / eB(z)$ is higher than the jet radius $r_{\mathrm{jet}}(z)$ (Hillas criterion). We calculated the maximum energies of each population of particles at different heights in the jet, dividing the acceleration region into ten sub-regions in which we assumed constant the magnetic energy and particle densities. In all cases, the maximum energies verify the Hillas criterion.\par
In the one-zone approach, the steady-state particle distribution $N(E,z)$ can be obtained as the solution to the transport equation:
\begin{equation}
\frac{\partial }{\partial E}\left[\left.\frac{\mathrm{d}E}{\mathrm{d}t}\right\vert_{\mathrm{loss}}N(E,z)\right] + \frac{N(E,z)}{t_\mathrm{esc}} = Q(E,z),
\label{equation_ParticleDistribution}
\end{equation}
where $t_\mathrm{esc} \approx \Delta z / v_{\mathrm{jet}}$ is the particle escape time from the acceleration region. The exact solution to Eq. (\ref{equation_ParticleDistribution}) is given by \citep[see][]{khangulyan2007}:
\begin{equation}
N(E,z)= \left\vert\frac{\mathrm{d}E}{\mathrm{d}t}\right\vert_{\mathrm{loss}}^{-1}\int_{E}^{E^{\mathrm{max}}(z)} \mathrm{d}E^{\prime}Q(E^{\prime},z)\exp\left(-\frac{\tau(E,E^{\prime})}{t_\mathrm{esc}}\right),
\label{equation_SolutionKhangaluyan}
\end{equation}
where
\begin{equation}
\tau(E,E^{\prime})=\int_{E}^{E^{\prime}} \mathrm{d}E^{\prime \prime} \left\vert\frac{\mathrm{d}E^{\prime \prime}}{\mathrm{d}t}\right\vert_{\mathrm{loss}}^{-1}.
\end{equation}
We calculate the spectral energy distribution of the radiation produced by primary and secondary particles in their interaction with magnetic, matter and radiation fields within the jet. Detailed formulas for the photon emissivity of each radiative process can be found in \cite{reynoso2009,romero2010a,romero2014,vieyro2012,vilatesis}; and references therein.\par
In the Table \ref{table_JetsParameters} we show the values of the different parameters of the inner jet.\par
\begin{table}[h!]
\caption{\small Parameters of the inner jet}             
\label{table_JetsParameters}      
\centering                          
\begin{tabular}{c c c c}        
\hline\hline                 
Parameter & Value \\    
\hline                        
   Jet luminosity [$L_{\mathrm{jet}}$]$^{\ddagger}$ & $10^{41}$ $\mathrm{erg}\,\mathrm{s}^{-1}$ \\
   Jet's bulk Lorentz factor [$\Gamma_{\mathrm{jet}}$]$^{\dagger}$ & $9$ \\
   Jet semi-opening angle tangent [$\chi$]$^{\dagger}$ & $0.1$  \\
   Jet's content of relativistic particles [$q_{\mathrm{rel}}$]$^{\dagger}$ & $0.1$ \\ 
   Hadron-to-lepton energy ratio [$a$]$^{\dagger}$ & $100$ \\
   Jet's launching point [$z_{\mathrm{0}}$]$^{\dagger}$ & $100$ $r_{\mathrm{g}}$ \\
   Magnetic field at base of jet [$B_{\mathrm{0}}$]$^{\ddagger}$ & $ 10^{7}$ $\mathrm{G}$ \\
   Cold matter density at $z_{\mathrm{0}}$ [$n_{\mathrm{0}}$]$^{\ddagger}$ & $6\times 10^{14}$ $\mathrm{cm}^{-3}$ \\
   Size of injection point [$\Delta z_{\mathrm{acc}}$]$^{\dagger}$ & $200$ $r_{\mathrm{g}}$ \\  
   Index for magnetic field [$m$]$^{\dagger}$ & $1.0$\\
\hline                                   
\
\small $^{\dagger}$ fixed \\
\small $^{\ddagger}$ calculated
\end{tabular}
\end{table}
\subsubsection*{Terminal jet}
We analyze the terminal region of the jet, considering two different epochs for the microquasar age $t_{\mathrm{MQ},1} = 10^{4}\,\mathrm{yrs}$ and $t_{\mathrm{MQ},2} = 10^{5}\,\mathrm{yrs}$. For the calculations we apply the model developed by \cite{bordas2009}. The distance to the compact object where the bow-shock occurs is:
\begin{equation}
l_{\mathrm{b}}(t_{\mathrm{MQ},1}) = \left(\frac{L_{\mathrm{jet}}}{\rho_{\mathrm{IGM}}}\right)^{1/5}t_{\mathrm{MQ},1}^{3/5},
\end{equation}
and
\begin{equation}
l_{\mathrm{b}}(t_{\mathrm{MQ},2}) = \left(\frac{L_{\mathrm{jet}}}{\rho_{\mathrm{IGM}}}\right)^{1/5}t_{\mathrm{MQ},2}^{3/5}.
\end{equation}
We adopt a baryonic density for the external medium at cosmological redshift $z = 10$ given by $\rho_{\mathrm{IGM}} \approx 2.4\times 10^{-4}\,\mathrm{cm^{-3}}$ \citep{douna2018}.\par
The speed of the bow-shock is given by:
\begin{equation}
v_{\mathrm{b}}(t) = \frac{\mathrm{d}}{\mathrm{d}t}l_{\mathrm{b}} = \frac{3}{5} \frac{l_{\mathrm{b}}}{t_{\mathrm{MQ}}},
\end{equation}
where it is reasonable to expect that $v_{\mathrm{b}}(t_{\mathrm{MQ},2}) < v_{\mathrm{b}}(t_{\mathrm{MQ},1})$, since for $10^{5}\,\mathrm{yrs}$ the jet will have traveled a longer distance in the intergalactic medium.\par
The pressure in the cocoon is approximately equal to that of in the shell, and is given by:
\begin{equation}
P_{\mathrm{c}}=P_{\mathrm{b}}= \frac{3}{4}\rho_{\mathrm{IGM}}v_{\mathrm{b}}^{2},
\end{equation}
where, as the age of the source increases, the inertia of the jet must decrease.\par
From the position of the recollimation shock the jet ceases to be conical and has a constant width. As we have assumed an opening angle given by $ \chi = \tan \theta_{\mathrm{jet}} = 0.1$ (see Table \ref{table_JetsParameters}), the position to the compact object where the recollimation-shock occurs is: 
\begin{equation}
z_{\mathrm{recoll}} \approx \sqrt{\frac{2\,L_{\mathrm{jet}}v_{\mathrm{jet}}}{(\gamma + 1)(\Gamma_{\mathrm{jet}}-1)\,\pi c^{2}P_{\mathrm{c}}}}.
\end{equation}
Here, we have assumed that the adiabatic index of the material in the cocoon is $\gamma = 5/3$, and that the shocks are weakly relativistic.\par
The adopted magnetic field in the downstream regions is such that the magnetic energy density is $\sim10\%$ of the thermal energy density of the gas \citep{bordas2009} In addition, in each acceleration region, the fraction of kinetic power converted into non-thermal particles is set at $1\%$ of the power of the jet. We have analyzed only the leptonic contribution here, since is the dominant one \citep{bordas2009}.\par
The injection function in the recollimation-shock, the reverse-shock, and the bow-shock, is assumed to be by $Q(E) = Q_{0}\, E^{-2}$, with the normalization constant $Q_{0}$ calculated as in Eq. (\ref{equation_LuminosityParticles}).\par
The cooling processes for non-thermal electrons in this region are: synchrotron radiation, inverse Compton scattering, and adiabatic losses. The target radiation field for inverse Compton losses is the cosmic microwave background. In the reconfinement region there are not adiabatic losses since the jet radius keeps constant. Because of the low density of matter in the cocoon and the reconfinement region, relativistic Bremsstrahlung radiation is negligible in these two zones and is only computed in the shell. \par
The model parameters for the three emitting zones in the terminal region of the jets that we adopt are listed in Table \ref{table:JetTerminal}.\par
\begin{table}[h!]
\caption{\small Parameter values adopted for the three emitting zones in the jet's terminal region
}             
\label{table:JetTerminal}      
\centering                          
\begin{tabular}{c c c}        
\hline\hline                 
Parameter & \qquad \qquad Value & \\    
\hline                        
   $n_{\mathrm{IGM}}$: IGM density [$\mathrm{cm}^{-3}$]$^{\dagger}$ & \qquad \qquad  $10^{-4}$ & \\
   $t_{\mathrm{MQ}}$: Source age [$\times10^{4}$ yrs]$^{\dagger}$ & $1$ & $10$\\
\hline
   \textit{SHELL} \\
\hline                        
   $B$: Magnetic field [$\times 10^{-5}\;\mathrm{G}$]$^{\ddagger}$ & $5.7$& $2.3$\\
   $v_{\mathrm{b}}$: Shock velocity [$\times 10^{8} \mathrm{cm}\,\mathrm{s}^{-1}$]$^{\ddagger}$ & $7.2$ & $2.9$\\
   $r$: Emitter size [$\times 10^{20}\;\mathrm{cm}$]$^{\ddagger}$ & $1.3$ & $5$\\
   $E_{\mathrm{max}}$: Max energy [$\mathrm{TeV}$]$^{\ddagger}$ & $3.1$ & $3.3$\\
   $n_{\mathrm{t}}$: Target density [$\times 10^{-4}\;\mathrm{cm}^{-3}$] & $9.6$ & $9.6$\\
\hline    
   \textit{COCOON} \\
\hline                        
   $B$: Magnetic field [$\times 10^{-4}\;\mathrm{G}$]$^{\ddagger}$ & $4.6$ & $1.9$\\
   $v_{\mathrm{sh}}$: Shock velocity [$\times 10^{10}\;\mathrm{cm}\,\mathrm{s}^{-1}$]$^{\dagger}$ & $2.98$ & $2.98$\\
   $r$: Emitter size [$\times 10^{18}\;\mathrm{cm}$]$^{\ddagger}$ & $2.5$ & $6.3$\\
   $E_{\mathrm{max}}$: Max energy [$\mathrm{PeV}$]$^{\ddagger}$ & $1.0$ & $1.0$\\
\hline  
   \textit{RECONFINEMENT} \\
\hline                        
   $B$: Magnetic field [$\times 10^{-3}\;\mathrm{G}$]$^{\ddagger}$ & $4.6$ & $1.9$\\
   $v_{\mathrm{conf}}$: Shock velocity [$\times 10^{9}\;\mathrm{cm}\,\mathrm{s}^{-1}$]$^{\dagger}$ & $2.98$ & $2.98$\\
   $r$: Emitter size [$\times 10^{20}\;\mathrm{cm}$]$^{\ddagger}$ & $3.8$ & $15$\\
   $E_{\mathrm{max}}$: Max energy [ $\mathrm{TeV}$]$^{\ddagger}$ & $33.4$ & $33.4$\\
\hline    
\small $^{\dagger}$ fixed \\
\small $^{\ddagger}$ calculated
\end{tabular}
\end{table}
\section{Results}
\label{sect:results}
In this paper we focus mainly on the calculation of the SED of the radiation emitted in the different components of the Population III microquasar that we have modeled, in order to start a quantitative description of its contribution to the reionization. Discussion of other physical properties of interest such as density, optical depth, or radial velocity distributions in the plasma of the accretion disk will be presented in a forthcoming work about super-Eddington sources and their outflows.\par
\subsection{Thermal-emission of the disk and wind}
The maximum temperature in the disk is approximately $10^{7}\,\mathrm{K}$, similar to the one estimated for the Galactic microquasar SS433 \citep[see][]{fabrika2007}. In addition, we do not report significant differences when the advection parameter varies. Once the radial distribution of effective temperature has been calculated, we can build the SED of the radiation produced by the accretion disk. Because of the high temperatures of the disk, it is expected that the highest contribution of electromagnetic emission occurs in the X-ray band. The bolometric luminosity of the accretion disk is approximately $10^{40}\,\mathrm{erg\,s^{-1}}$, very similar to that estimated for SS433 \citep[see][]{fabrika2007}.\par
\begin{figure*}[h!]
   \centering
   \includegraphics[scale=1.2]{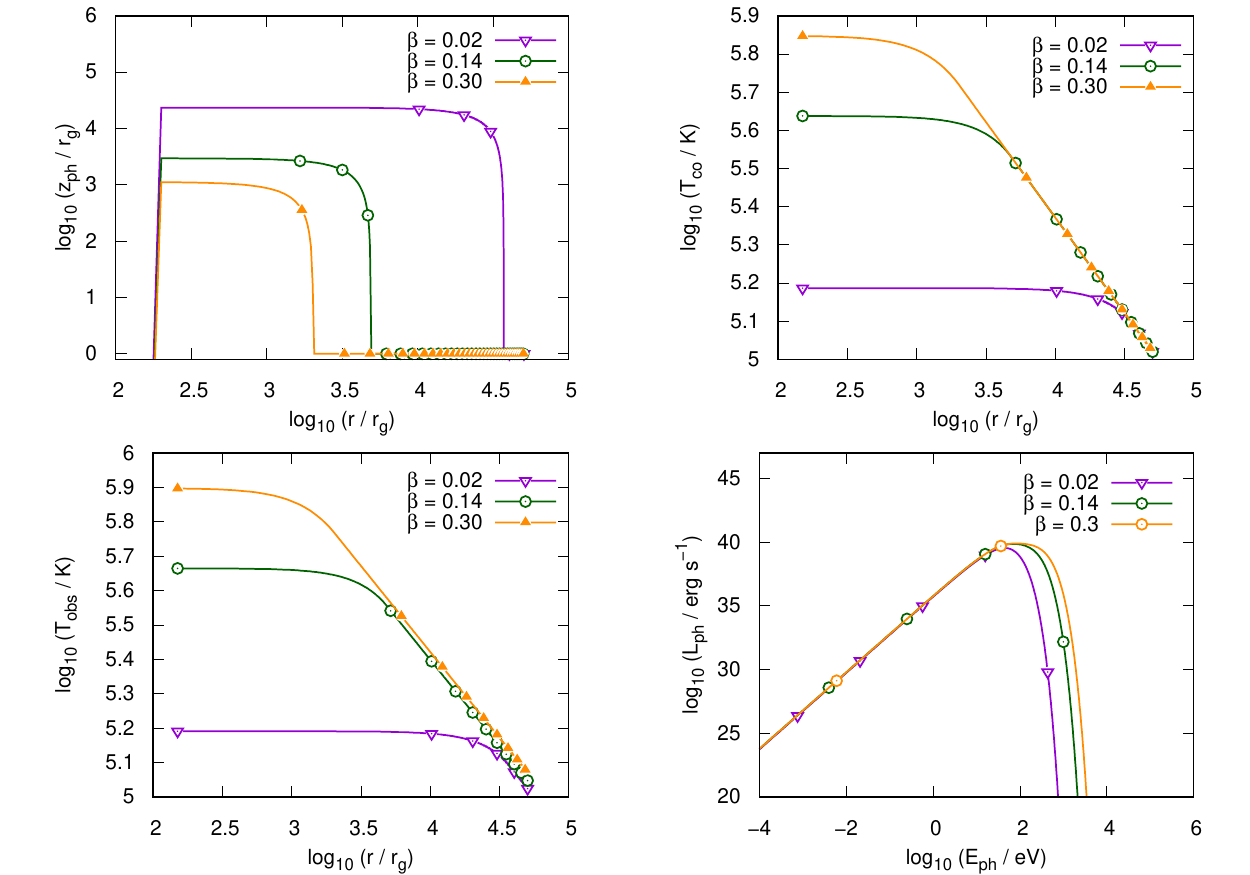}
	   \caption{\small Radial distribution of the height, comoving temperature, and observed temperature of the gas at the wind apparent photosphere, and spectral energy  distribution of the radiation produced at the wind apparent photosphere measured by an observer at infinity. We consider three values for the wind velocity $\beta = v/c$ in all cases.}
     \label{fig:properties_wind}
\end{figure*}

Figure \ref{fig:properties_wind} shows several physical quantities of the radiation-pressure driven wind for the values of the adopted parameters in this model. We consider three velocity values for the wind that correspond to the escape velocity in $r_{\mathrm{cr}}$ ($\beta = 0.02$), in $r_{0}$ ($\beta = 0.14$), and a velocity considerably higher than the previous ones ($\beta = 0.30$). We present the radial distribution of the height, comoving temperature and observed temperature at the wind apparent photosphere, and the spectral energy distribution of the radiation produced by the wind at the apparent photosphere. Considering that the wind speed is constant, we analytically calculate the height of the apparent photosphere \citep[see][]{fukue2010}.\par 
The apparent photosphere decreases rapidly as the wind speed increases. This is because for higher values of wind speed, the density of the gas is reduced. The temperature of the gas varies between $T = 10^{5} - 10^{6}\,\mathrm{K}$ in comoving and inertial reference frame. Because the wind speed is generally small compared to the speed of light, there are no significant variations with the viewing angle (i.e., the Doppler factor is approximately $\sim 1$).\par
The spectrum of the radiation emitted by the gas of the wind is that of a multi-temperature blackbody. In general, the main contribution of the wind to the SED measured by an observer at infinity is in the range of ultraviolet to soft X-rays ($E_{\mathrm{ph,max}} \sim 0.1 - 1\,\mathrm{KeV}$). Again, there are no significant differences when varying the viewing angle.\par
\begin{figure}[h!]
\centering
   \includegraphics[width=0.5\textwidth]{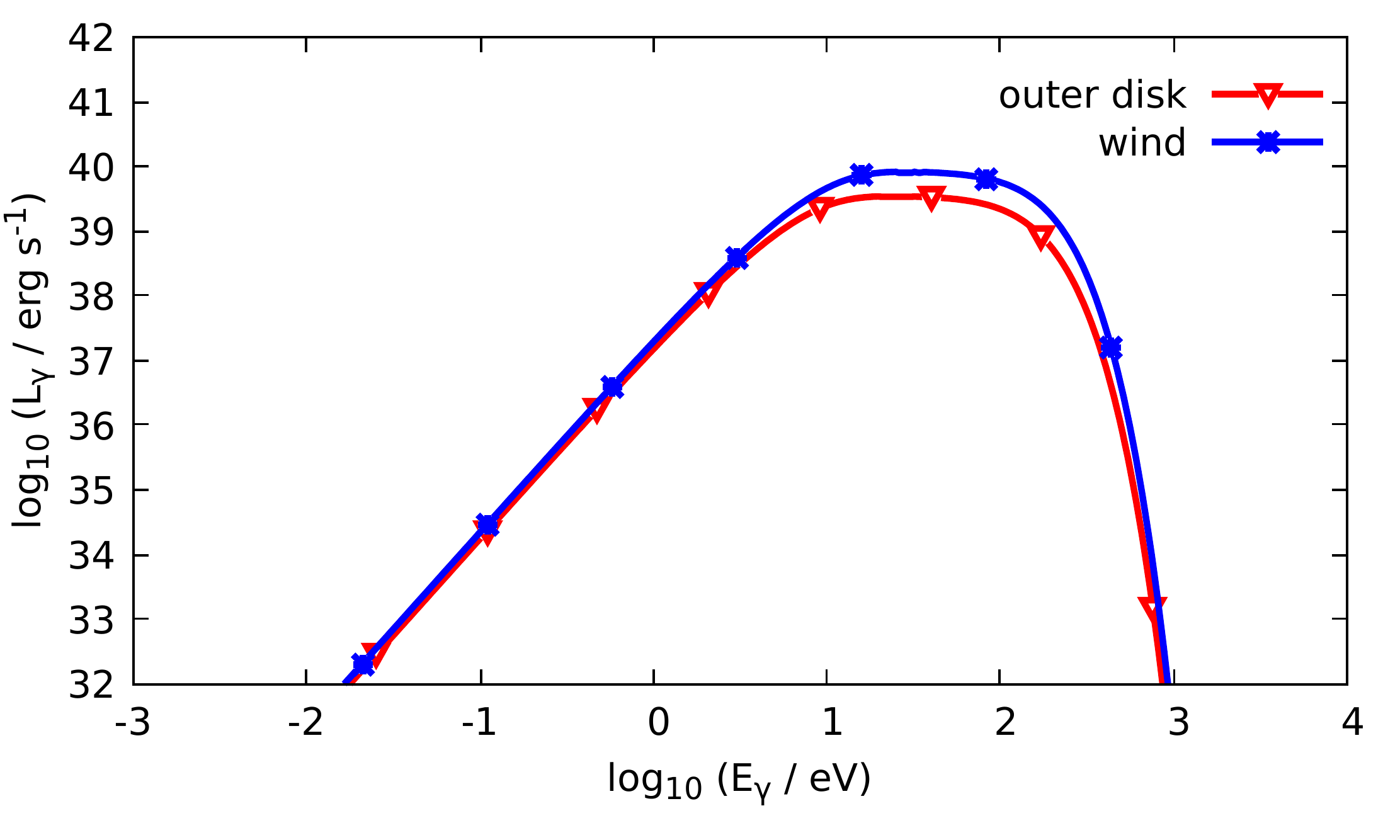}\\
  \caption{\small Comparation of the thermal SED of the radiation produced by the disk and wind. On the outer disk the emission is not hidden by the wind. This happens for $r > r_{\rm cr} \approx 3600\,r_{\rm g}$. We consider for the disk that the advection parameter is $f=0.5$. For the wind, we assume a constant escape velocity given by $v = 0.14c$. The viewing angle is $\theta$ = 60\grad.}
  \label{fig:SED_accretion_disk2}
\end{figure}

In Fig. \ref{fig:SED_accretion_disk2} we show a comparison between the thermal SEDs of the wind and the disk. We show only the SED of the disk that is not hidden by the wind, i.e. from what we call the outer disk. The emission is mainly in the ultraviolet band, and determined by the wind. The radiation emitted in the innermost region of the disk will attenuate when interacting with the wind particles. This last leads to an enhancement in the energy of the photons of the wind. Therefore, the UV emission of the wind photosphere that we have calculated can increase to $E\sim 1\,{\rm keV}$. For a detailed description of the interaction between disk radiation and wind particles, numerical simulations should be implemented.\par
\subsection{Non-thermal emission of the jet}
\subsubsection*{Inner jet}
In our jet model there are several free parameters that determine different scenarios for non-thermal emission and cosmic ray production. Our model is defined by the choice of the following parameters: the minimum energy of the accelerated particles is $E_{\mathrm{min}} = 2\,mc^{2}$, the particle acceleration mechanism is efficient with $\eta = 0.1$, the relativistic particle content of the jet is essentially hadronic $a = L_{\mathrm{p}}/L_{\mathrm{e}} = 100$, the particle acceleration region is close to the compact object $z_{\mathrm{acc}} = 300\,r_{\mathrm{g}}$, and the spectral index of the injection function is $p=2$. This set of parameters corresponds to the scenario that we denote as a), and will serve to compare the effect on the SED when we vary these parameters one by one. We consider the cases in which b) the minimum energy of the accelerated particles is $E_{\mathrm{min}} = 100\,mc^{2}$, c) the particle acceleration mechanism is inefficient $\eta = 10^{-4}$, d) there is equipartition between the leptonic and hadronic content of relativistic particles in the jet $a = L_{\mathrm{p}}/L_{\mathrm{e}} = 1$, e) the acceleration of particles occurs at a distance from the compact object similar to the orbital semi-axis of the binary system $z_{\mathrm{acc}} = 5\times 10^{5}\,r_{\mathrm{g}}$, and f) the spectral index of the injection function is hard ($p=1.5$). We assume for the jet that the size of the acceleration region is $\Delta z_{\mathrm{acc}} = 200\,r_{\rm g}$, and the opening angle is 5\grad. We calculate that the magnetic field at the launch point of the jet is $B = 10^{7}\,{\rm G}$ in all cases. \par
\begin{figure*}[h!]
 \centering
    \subfloat[\small ]{
      \label{fig:SED_comparation1}
       \includegraphics[scale=0.55]{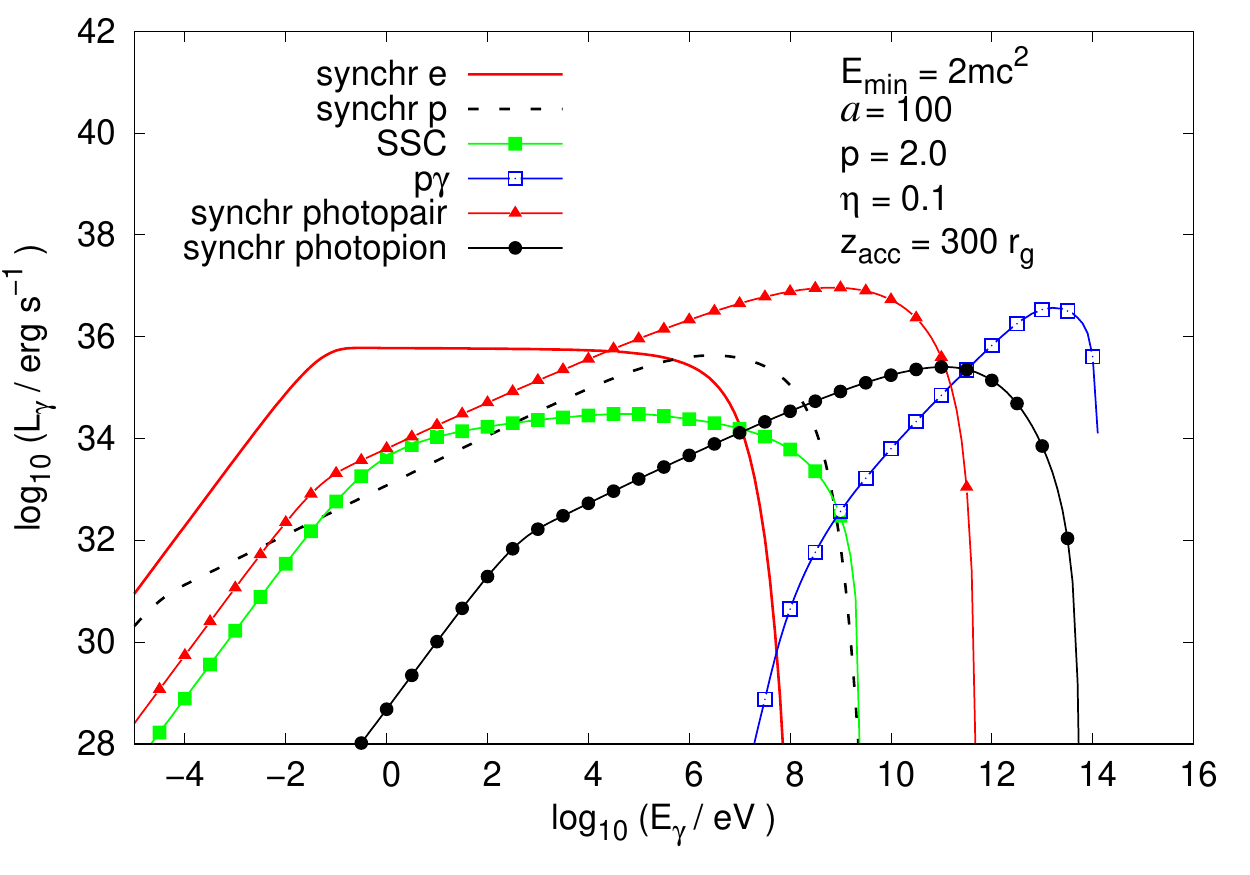}}
       \subfloat[\small ]{
      \label{fig:SED_comparation2}
      \includegraphics[scale=0.55]{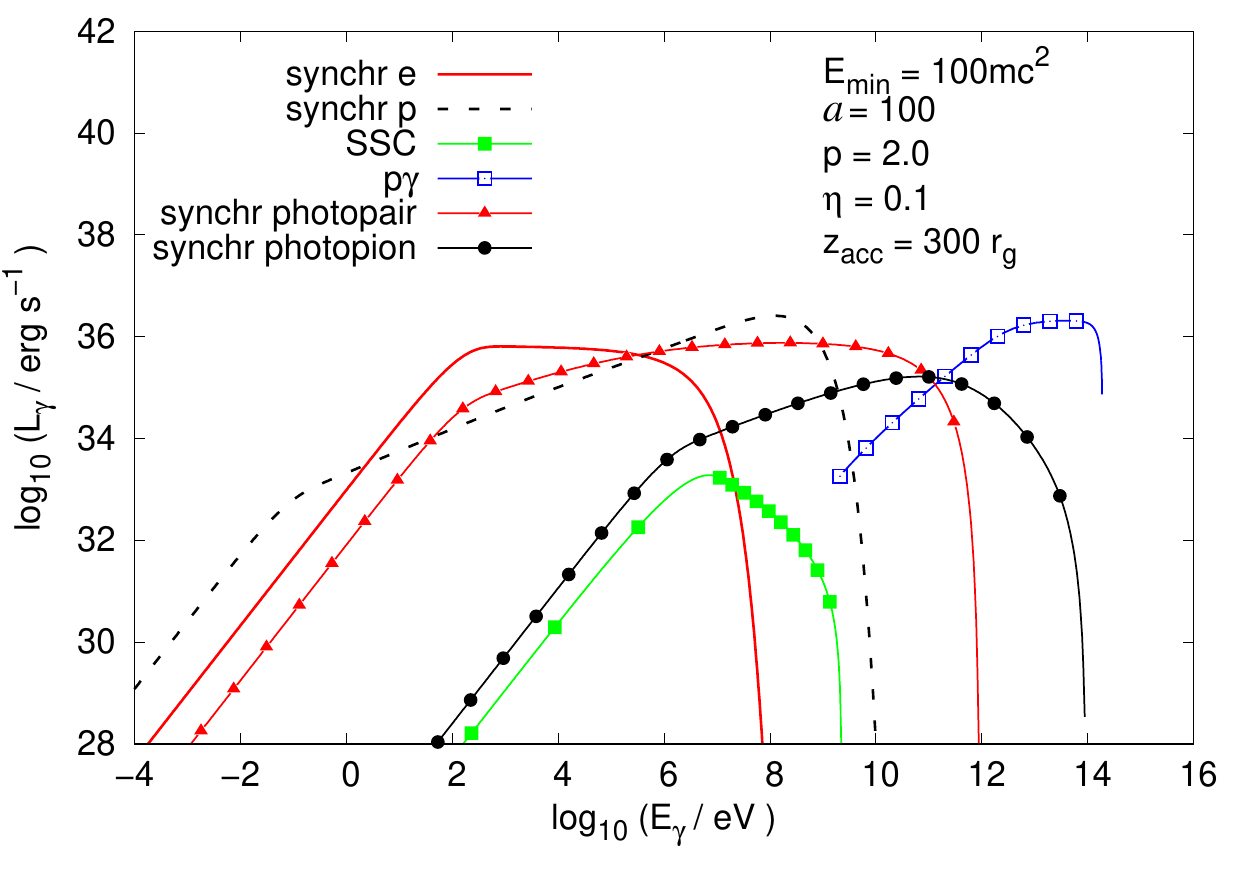}}\\
          \subfloat[\small ]{
      \label{fig:SED_comparation3}
       \includegraphics[scale=0.55]{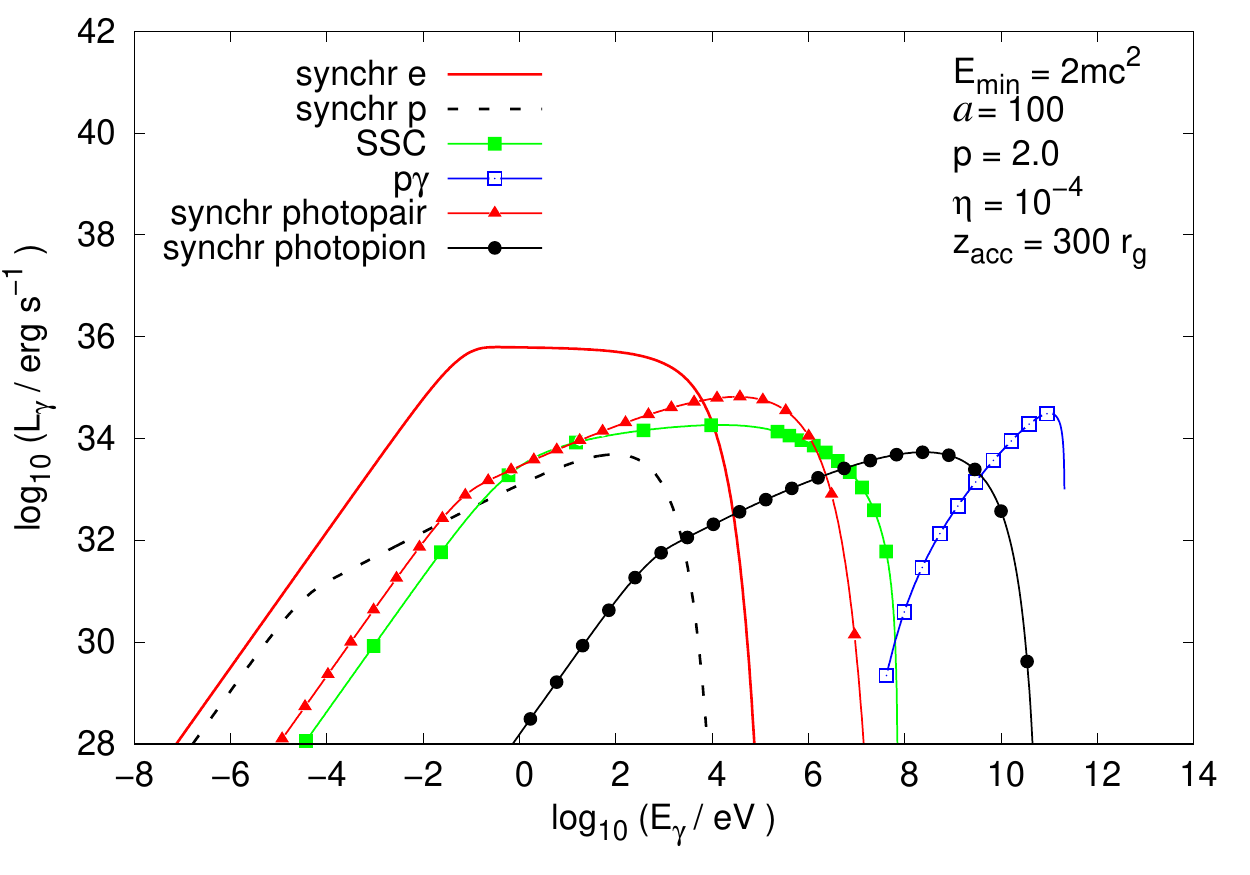}}
       \subfloat[\small ]{
      \label{fig:SED_comparation4}
      \includegraphics[scale=0.55]{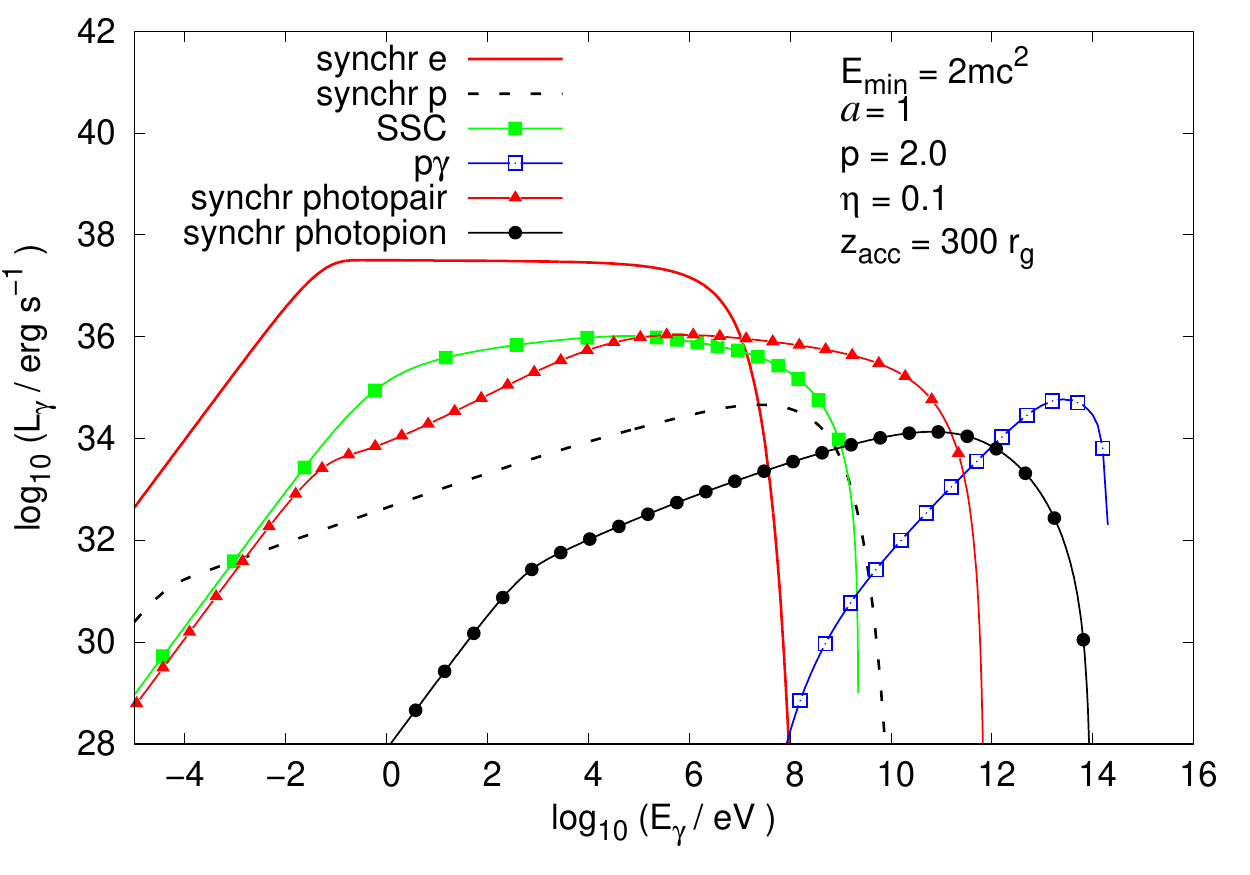}}\\
          \subfloat[\small ]{
      \label{fig:SED_comparation5}
       \includegraphics[scale=0.55]{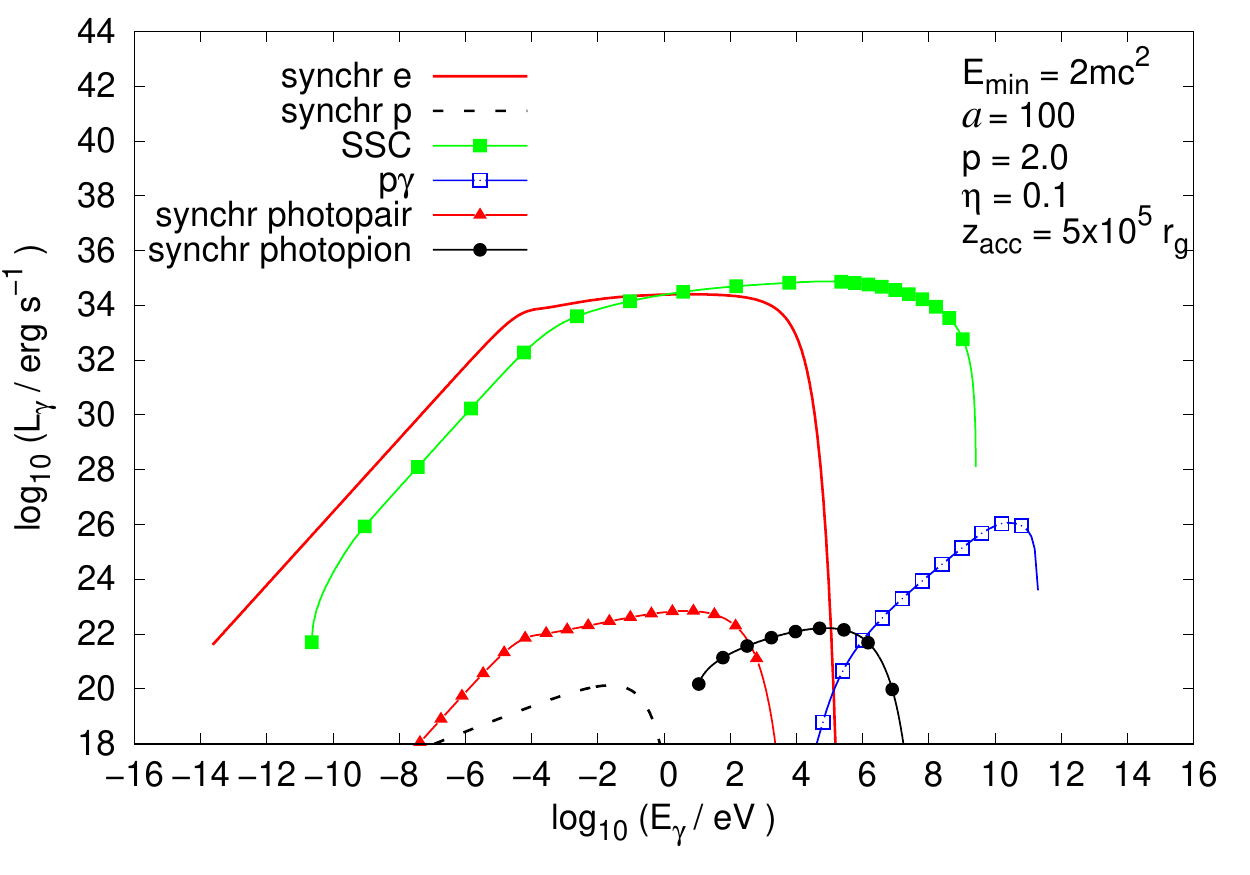}}
       \subfloat[\small ]{
      \label{fig:SED_comparation6}
      \includegraphics[scale=0.55]{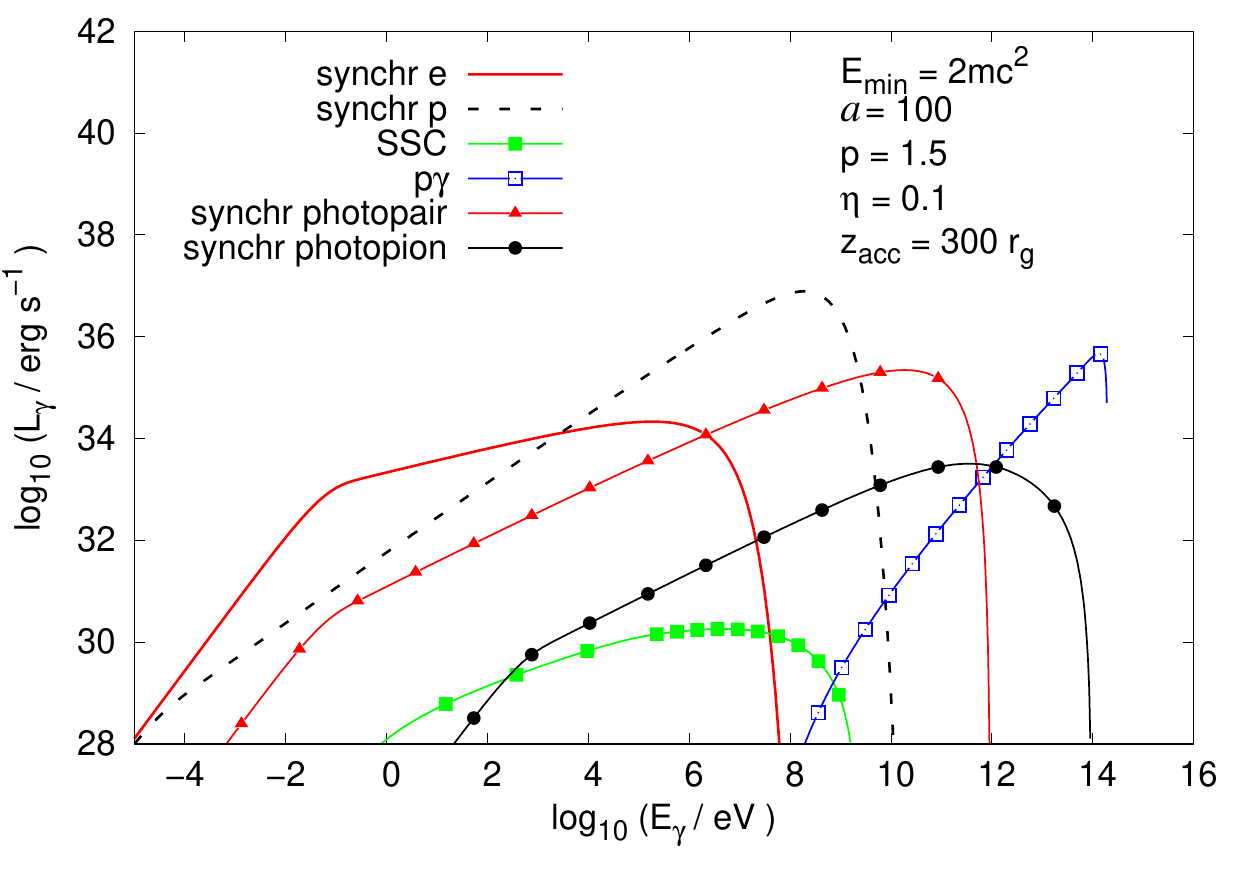}}
     \caption{\small Most relevant individual contributions for the total SED of the inner jet without considering absorption effects for different choices of sets of parameters. In all cases the adopted viewing angle is $\theta$ = 60\grad.}
      \label{fig:comparation SEDs}
\end{figure*}

Figure \ref{fig:comparation SEDs} shows the non-thermal spectral energy distributions of the  radiation emitted in the jet by the population of primary and secondary particles for different values of the model parameters. The SEDs are corrected by Doppler effect considering a viewing angle of $\theta =$ 60 \grad.\par
The main contributions to the SEDs vary according to the adopted parameters. The electron synchrotron radiation is the most important one when the leptonic content of the jet is greater than $a = 1$ and the acceleration of particles is inefficient $\eta = 10^{-4}$, peaking at $\approx 10^{37}\,\mathrm{erg\,s^{-1}}$ for $E_{\gamma}\sim 1 - 10^{6}\, \mathrm{eV}$. In the case where the acceleration region is at $z_{\mathrm{acc}} = 5\times10^{5}\,r_{\mathrm{g}}$ the SED is determined by the leptonic radiative processes, regardless the content of the jet. This is because the protons are advected outside the acceleration region without losing their energy, $t_{\mathrm{conv}} = 3.4\times10^{-3}\,\mathrm{s}$ and $t_{p\gamma} = 741\,\mathrm{s}$ for $E_{\mathrm{p}} = 1\,\mathrm{TeV}$, where the photohadronic collisions are the main cooling mechanism. In this latter scenario, the gamma emission is due to the inverse Compton scattering of primary electrons with synchrotron photons of electronic origin.
\par
The efficiency of the acceleration determines the maximum energy of the primary particles, and therefore the maximum energy of the photons i.e. the high-energy cutoff the SEDs. On the other hand, the minimum energy of the particles determines the low-energy cutoff of the SEDs.
\par
The contribution of synchrotron radiation from Bethe-Heitler pairs (synchrotron of photopairs) is in general relevant. This is a peculiar feature of this type of microquasar. It is the dominant radiative process for $E_{\gamma} \sim 10\,\mathrm{GeV}$, in the case of efficient acceleration, and for $E_{\gamma} \sim 100\,\mathrm{keV}$, in the case of inefficient acceleration. Synchrotron radiation of pairs produced by decay of charged pions in $p\gamma$ interactions is a relevant process for $E_{\gamma} \sim 10^{12}\,\mathrm{TeV}$, in the case of efficient acceleration, peaking at $L_{\gamma}\sim 10^{34 - 35}\,\mathrm{erg\,s^{-1}}$. The $p\gamma$ radiation is the most important contribution to the SEDs at very high energy, $E_{\gamma}\sim 10^{14}\,\mathrm{eV}$.
\par
If the size of the acceleration region is much higher, the main consequence is that the convection is less relevant, therefore the maximum energy of the protons becomes determined by the radiative losses.\par
\subsubsection*{Absorption of the high-energy emission of the inner jet}
\label{sect:absorption}
The attenuation of the radiation produced in the cooling region near the compact object by $\gamma \gamma$ annihilation with the stellar radiation field depends on the location of the particle acceleration region, the orbital configuration of the binary system, and the energy of the incident $\gamma$-rays \citep[see][]{romero2010b,dubus2006,reynoso2008a}. We report that when the compact object is in opposition the emission is partially suppressed, regardless of the choice of jet parameters.\par
\begin{figure}[h!]
  \centering
  \includegraphics[scale=0.7]{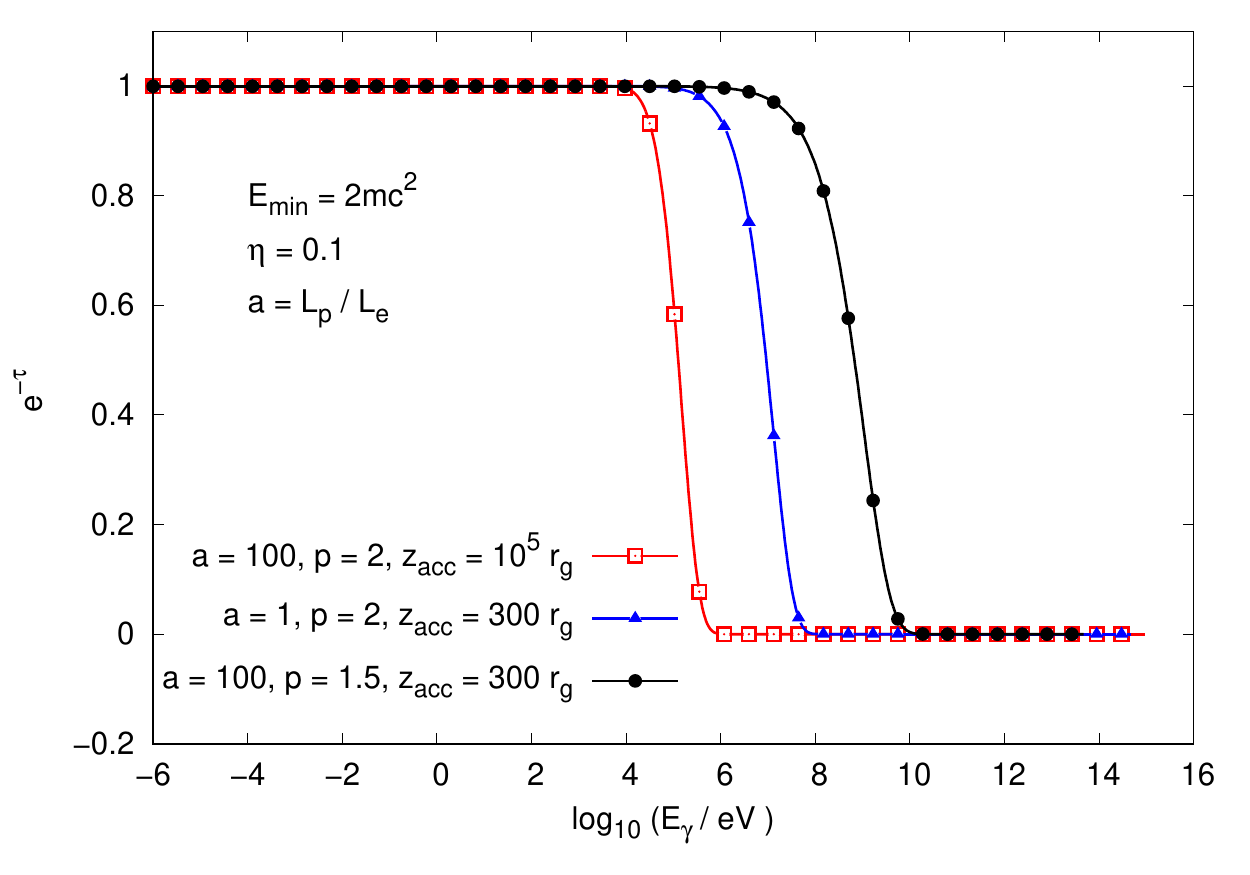}
     \caption{\small Attenuation curves for different models of inner jet.}
        \label{fig: absorption_comparation}
  \end{figure}

High-energy radiation emitted in the jets is also attenuated by $\gamma \gamma$-annihilation with synchrotron photons of the primary electrons produced internally. The radiation produced in the jet is completely suppressed for $E_{\gamma} > 100\,\mathrm{MeV}$ regardless of the choice of parameters adopted, as seen in Figure \ref{fig: absorption_comparation}. In the case where the acceleration of particles takes place at $z_{\mathrm{acc}} = 5\times10^{5}\,r_{\mathrm{g}}$ the radiation of highest energy that leaves the jet corresponds to hard X-rays $E_{\gamma}\sim 10\,\mathrm{keV}$. This absorbed radiation produces electron-positron pairs which are created with very high energies and are cooled mainly by synchrotron radiation, whereby the energy of the initial photons is distributed in synchrotron photons of lower energy, with no electromagnetic cascades taking place, since they are suppressed by the high magnetic fields.\par
\subsubsection*{Terminal jet}
\label{subsection:results of terminal jets}
\begin{figure}[h!]
\centering
  \begin{minipage}{0.6\textwidth}
    \includegraphics[width=0.7\textwidth]{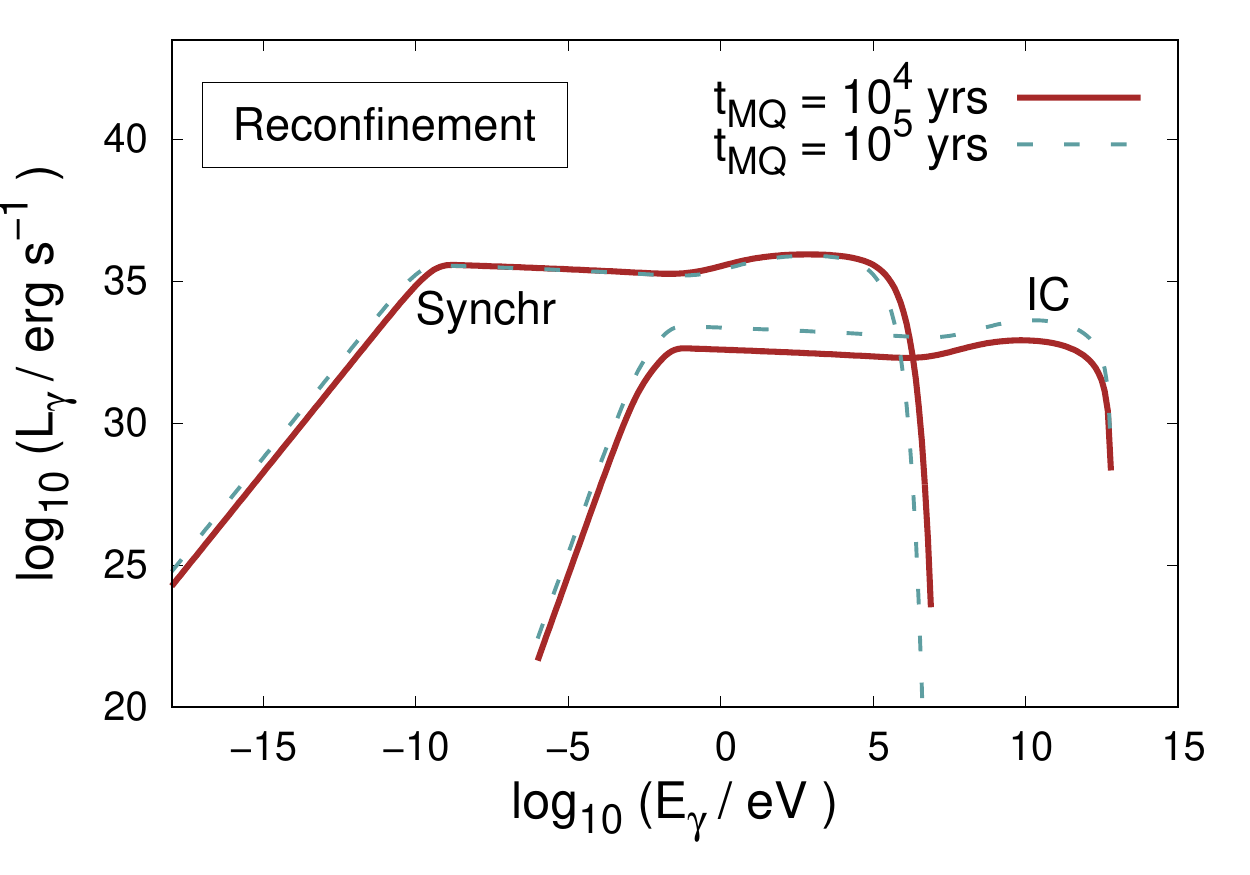}\\
   \end{minipage}%
  \hspace{5mm}
  \begin{minipage}{0.6\textwidth}
    \includegraphics[width=0.7\textwidth]{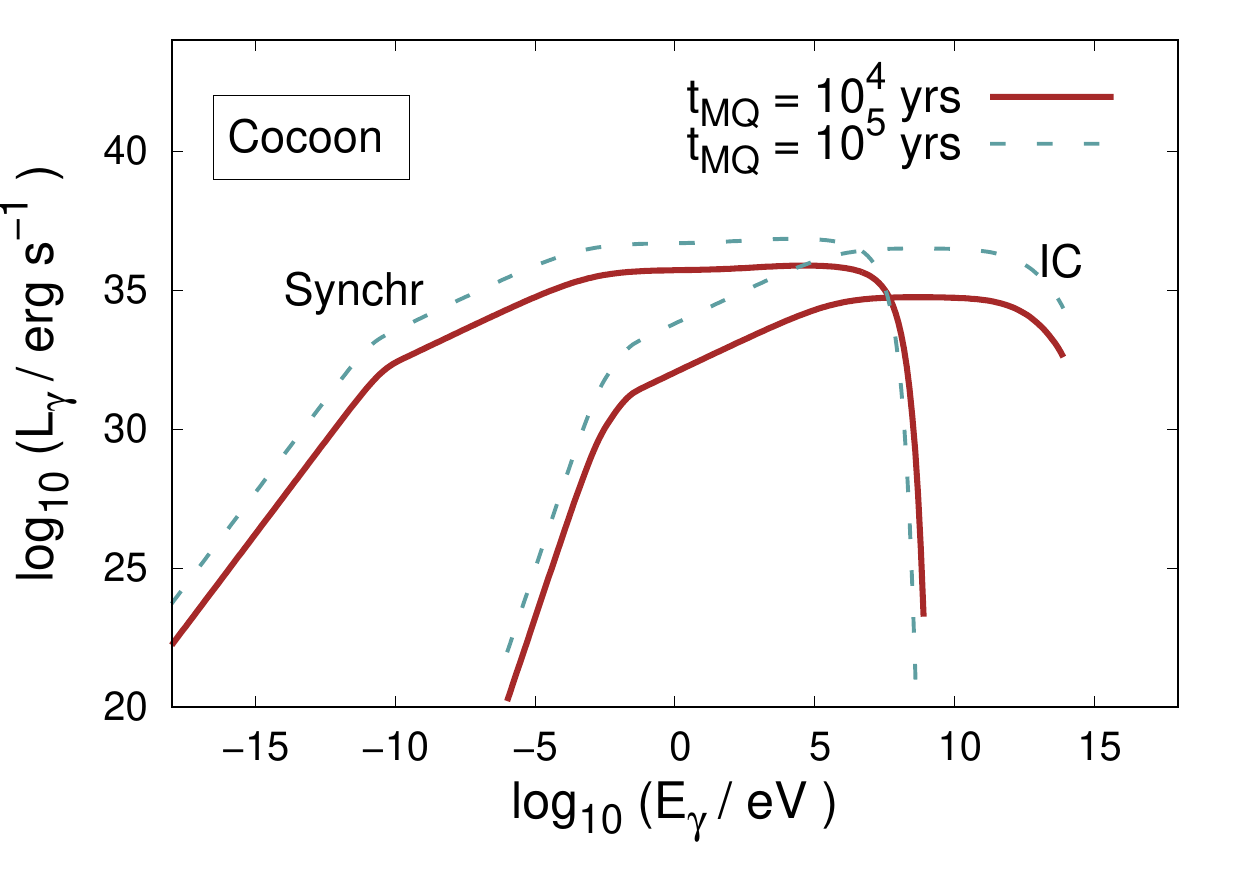}\\
  \end{minipage}
  \hspace{5mm}
  \begin{minipage}{0.6\textwidth}
    \includegraphics[width=0.7\textwidth]{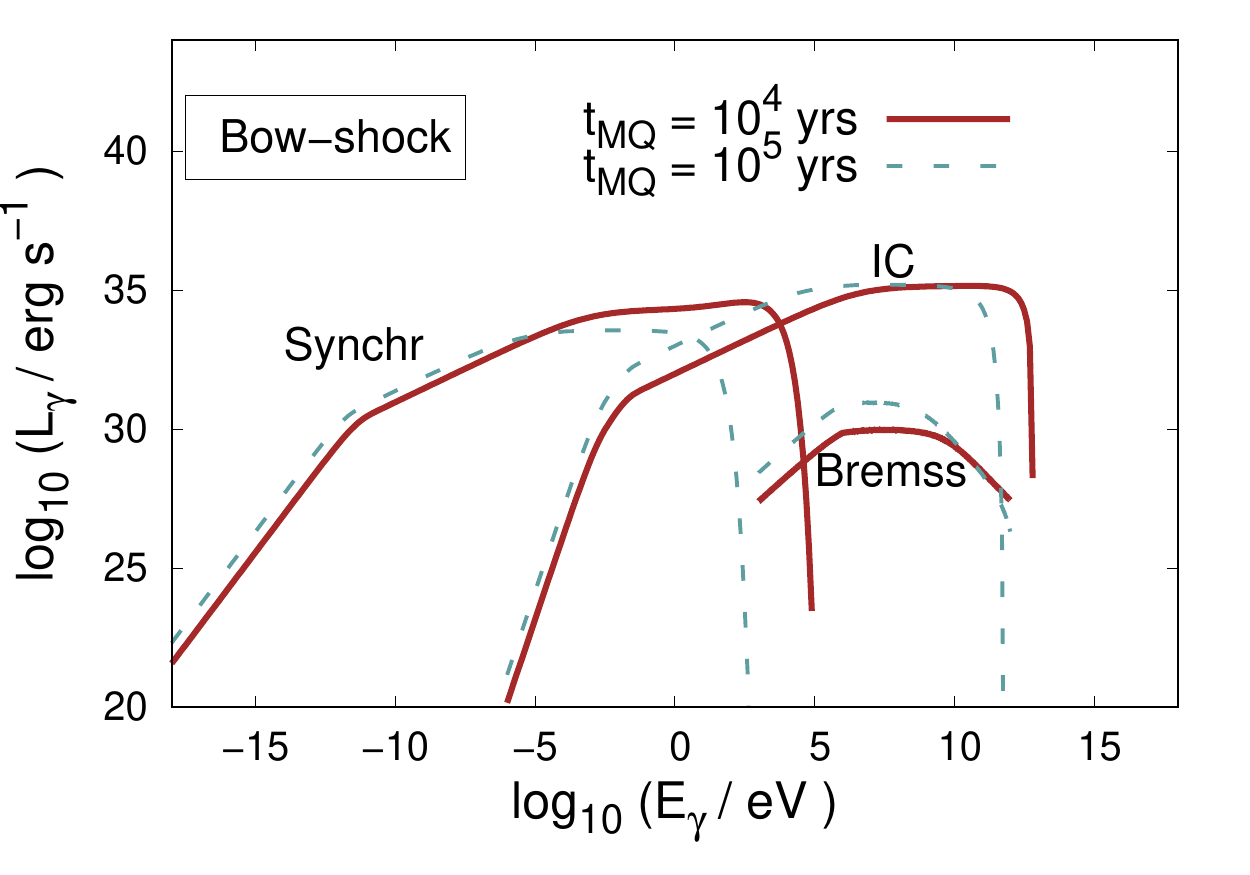}\\
  \end{minipage} 
  \caption{\small Non-thermal spectral energy distributions of the radiation produced in the reconfinement region, cocoon and bow-shock. Two epochs are considered for the source.}
  \label{fig:SED_terminaljet}
\end{figure}

In Fig. \ref{fig:SED_terminaljet} we show the SED of the radiation produced in the reconfinement region, cocoon, and the bow-shock by the relativistic electrons, for the two epochs of interest. In the region of reconfinement, the maximum energies of electrons for the different epochs are approximately 30 TeV, being always limited by synchrotron losses. Synchrotron emission is also the dominant radiative process.  In the cocoon, the shape of the non-thermal spectral energy distribution differs from that seen in the reconfinement region. This is because in the cocoon the cooling mechanism of the electrons varies according to the energy they have: for energies less than $10\,\mathrm{GeV}$ the main cooling mechanism is adiabatic losses, for energies between $10\, \mathrm{GeV}$ and $10\,\mathrm{TeV}$ the main mechanism is inverse Compton scattering, finally for higher energies the mechanism is synchrotron radiation. On the contrary, in the reconfinement region, the radiative cooling mechanism in the whole range of energies is synchrotron radiation. In the cocoon, the two contributions to the bolometric luminosity are comparable, in both epochs of the source. In the bow-shock, the luminosity produced by inverse Compton scattering dominates the emission.  We have included the spectral energy distribution by relativistic Bremsstrahlung, which turns out to be irrelevant because of the low density of the medium. This differs from the results obtained in known microquasars, where the emission by relativistic Bremsstrahlung could dominate the spectrum at high energies. In the same sense, the radiation produced by $pp$ collisions is irrelevant. So, a leptonic model is appropriate to describe the proposed scenario.
\par
\subsection{Total radiative output}
\label{sect:total_output}
\begin{figure}[h!]
\centering
  \begin{minipage}{0.6\textwidth}
    \includegraphics[width=0.7\textwidth]{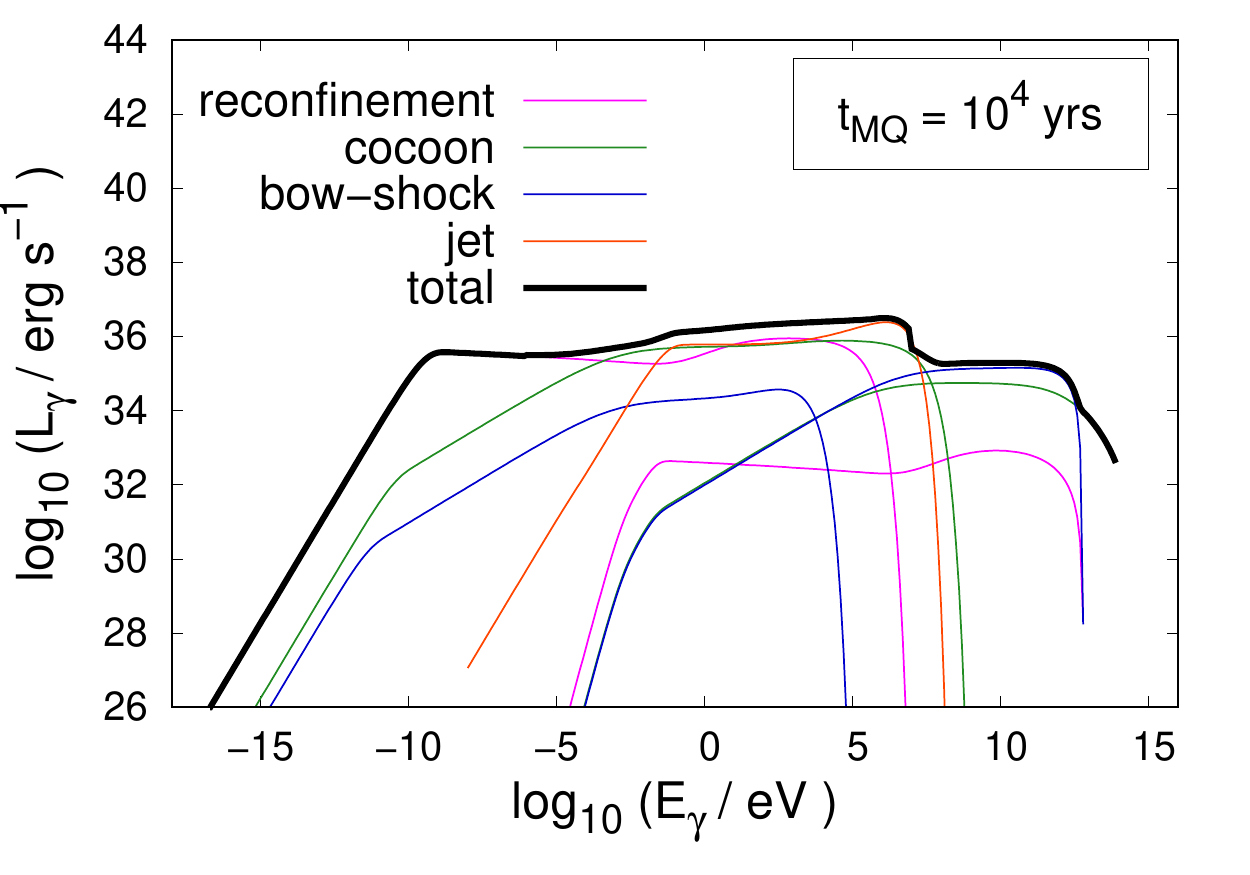}\\
  \end{minipage}%
  \hspace{5mm}
  \begin{minipage}{0.6\textwidth}
    \includegraphics[width=0.7\textwidth]{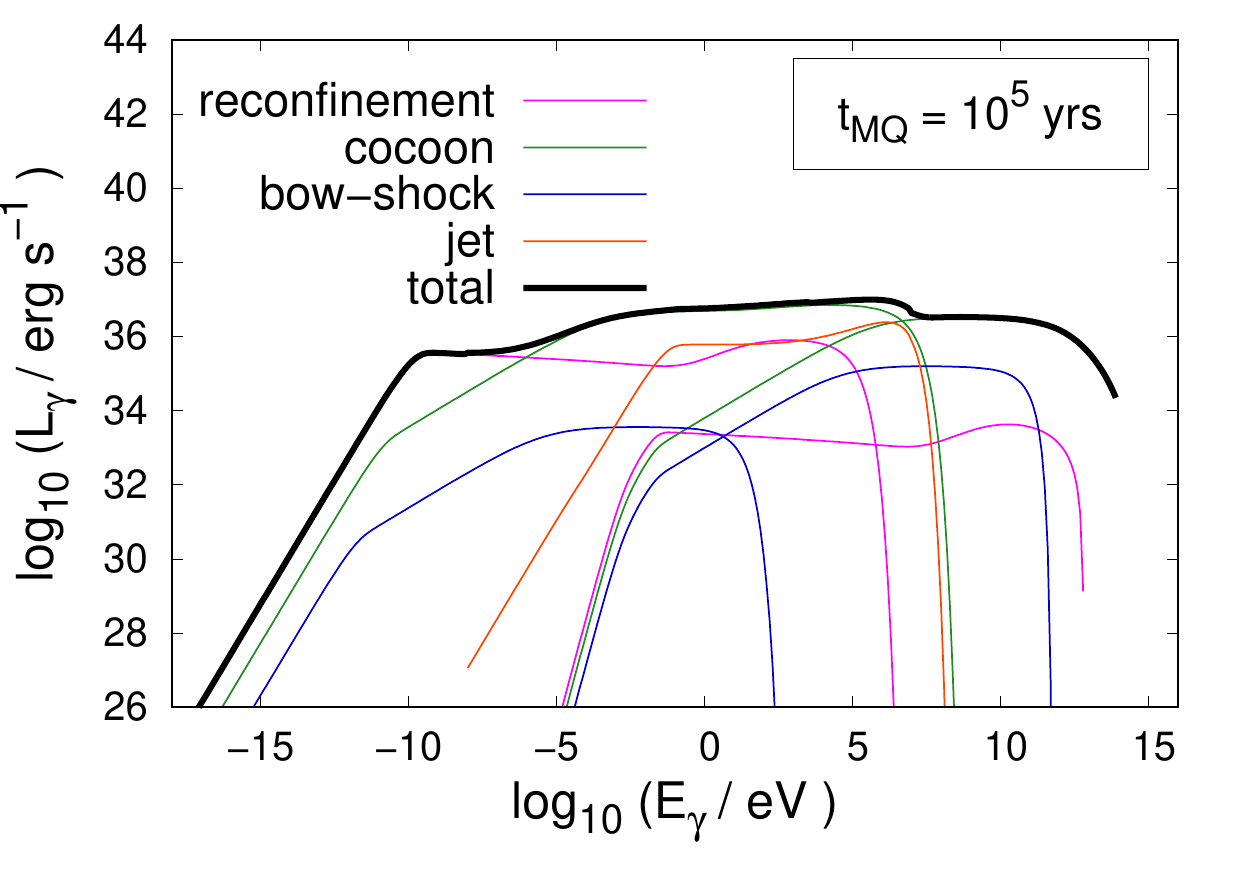}\\
  \end{minipage}
  \hspace{5mm}
  \begin{minipage}{0.6\textwidth}
    \includegraphics[width=0.7\textwidth]{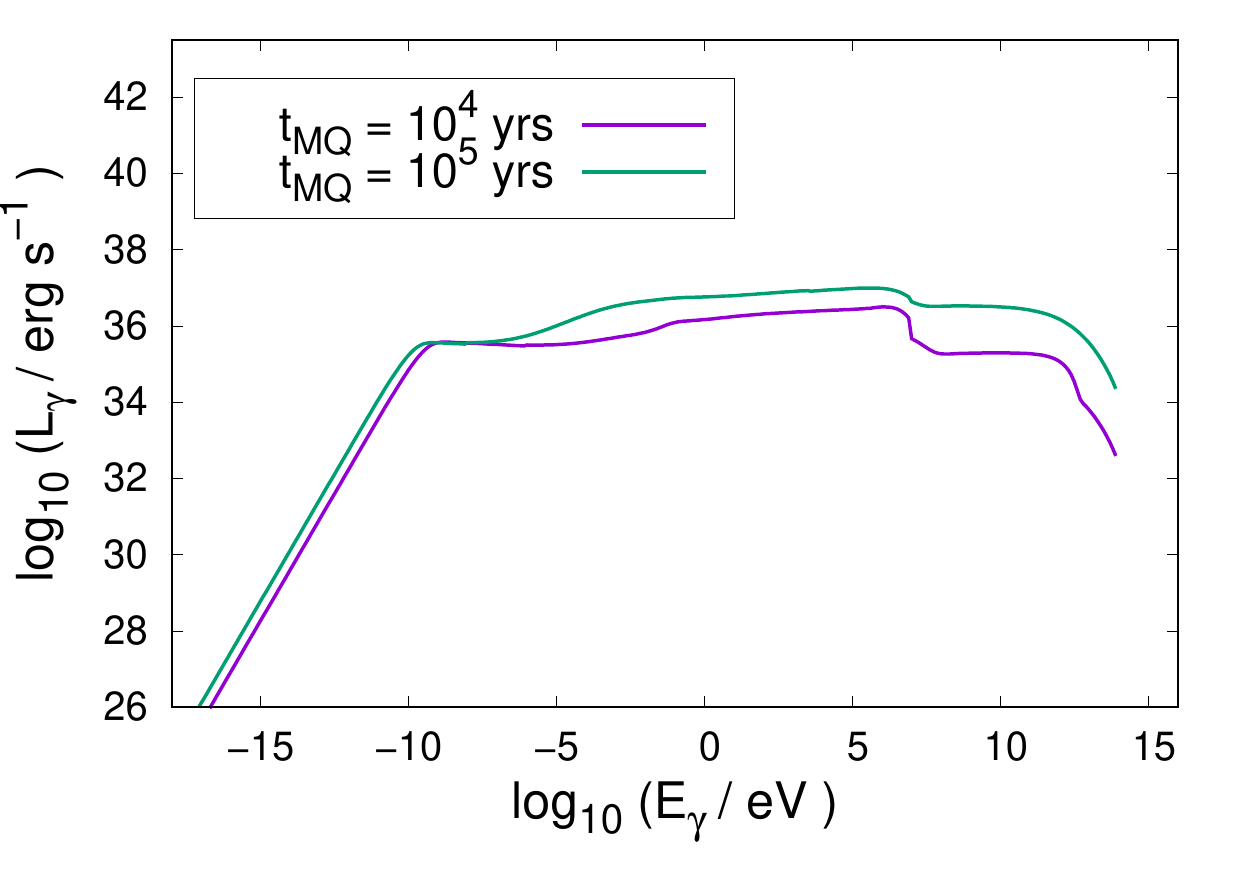}\\
  \end{minipage} 
  \caption{\small Total non-thermal spectral energy distribution for the epochs $t_{\mathrm{MQ}} = 10^{4}$ and $t_{\mathrm{MQ}} = 10^{5}$ years. The adopted viewing angle is $\theta$ = 60\grad.}
  \label{fig:totalSED}
\end{figure}

In Fig. \ref{fig:totalSED} we show the total non-thermal SEDs produced in the four regions of particle acceleration that have been considered for the epochs $t_{\mathrm{MQ}} = 10^{4}$ and $t_{\mathrm{MQ}} = 10^{5}$ years. We consider case a) for the inner jet model. In energies corresponding to the band of radio, infrared and optical, the emission is mainly due to the radiation produced in the reconfinement region and in the cocoon. For energies in the range of ultraviolet, X-rays and soft $\gamma$-rays, the radiation is predominantly emitted in the inner jet. For energies higher than 1 MeV, where the emission of the inner jet is self-absorbed, the predominant contribution is the inverse Compton radiation produced in the bow-shock. The electromagnetic emission reaches energies up to $E_{\gamma} \sim 1$ TeV. In addition, we show a comparison between the total non-thermal spectral energy distribution for the epochs $t_{\mathrm{MQ}} = 10^{4}$ years and $t_{\mathrm{MQ}} = 10^{5}$ years in which no significant differences are observed. This is partly caused by the constant accretion rate. In a realistic case, where the initial accretion rate is extremely super-Eddington and then decreases, differences are expected between the total non-thermal spectra for different epochs of the source.\par
\begin{figure}
\centering 
 \includegraphics[scale=.72]{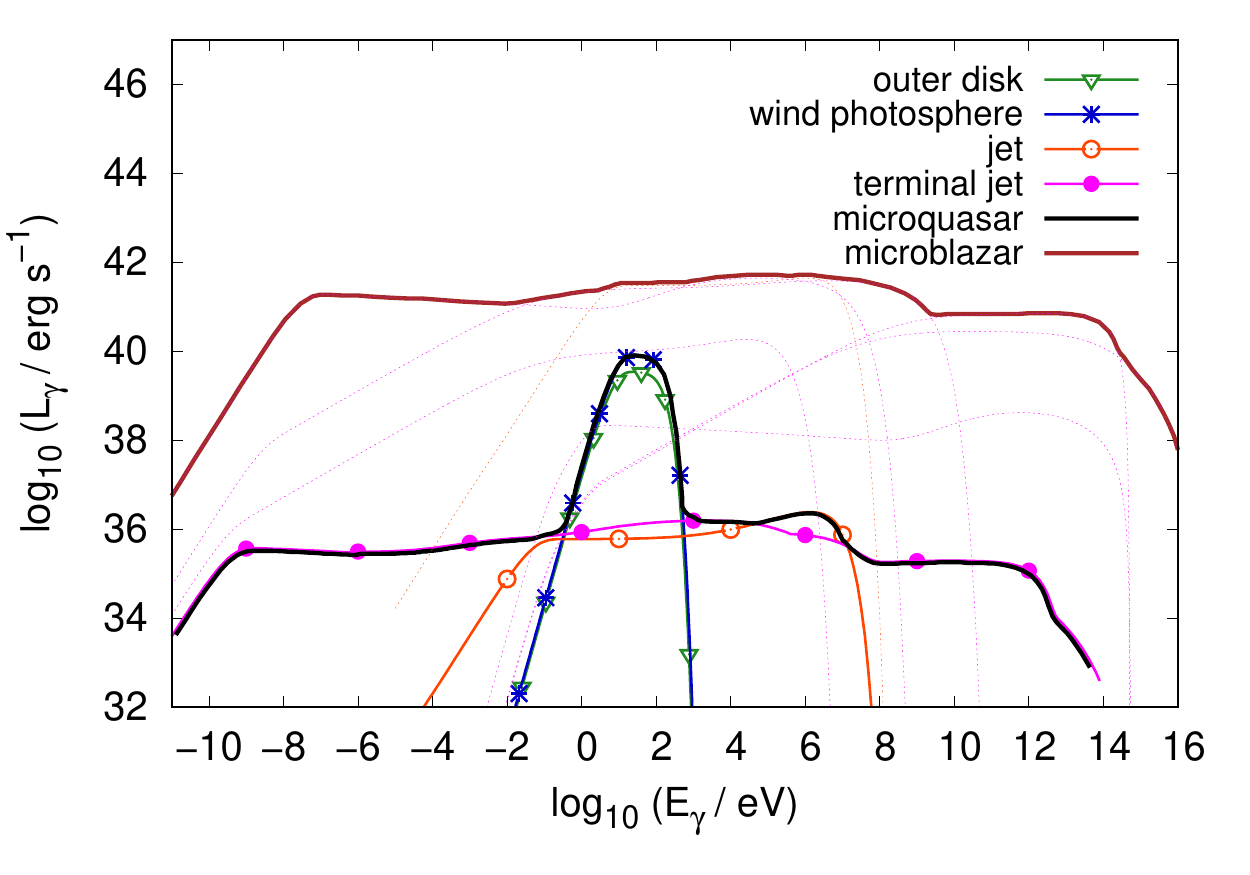}
 \caption{Total spectral energy distribution of the system. The inner disk is obscured by the wind and the most energetic photons of the disk are attenuated. We consider a viewing angle $\theta$ = 60 \grad. In addition, we show the SED in the case of a microblazar, $\theta$ = 1 \grad, where the radiation emitted in the jet and the terminal jet dominates the spectrum.}
 \label{fig:SEDtotal}
\end{figure}

In Fig. \ref{fig:SEDtotal} we show the total SED of the microquasar (outer accretion disk + wind photosphere + inner jet + terminal jet). Microquasars of Population III emit broadband radiation throughout the whole electromagnetic spectrum. The inner disk is obscured by the wind. The relative contribution of the radiation emitted in the jet to the total SED depends strongly on the viewing angle. We show the SED as observed at two different viewing angles: $60$\grad  (microquasar) and $1$\grad  (microblazar). In microquasars, for energies in the range of UV to soft X-rays, the spectrum is dominated by the thermal emission of the wind and the disk. The highest emission is because of the UV radiation of the wind. In microblazars, the inner jet and the terminal jet dominates the spectrum. 
It is important to emphasize that, in our model, all microquasars are microblazars seen from some reference frame. In the  case of the microblazars, although there is higher reionization in the direction to the observer, the phenomenon is the result of the intrinsic anisotropy, so its consideration for total reionization is irrelevant.
\section{Reionization by the first microquasars}
\label{sect: reionization}
In the previous sections we have determined that the first microquasars are powerful sources of thermal ultraviolet radiation by the wind, and broadband non-thermal radiation by the jets. In the following, we study their contribution to reionization through some specific mechanisms. We implement elementary calculations of the size of the ionized zone by UV photons of the wind, the production of relativistic protons in the environment of the inner jet by the neutron decay, and the role of accelerated particles in the terminal jet. A detailed treatment of these processes requires the implementation of numerical simulations. Electromagnetic cascades initiated by gamma rays of the jets can produce ionizing ultraviolet radiation and have a non-negligible contribution in reionization. This process will be developed in a forthcoming comunication.\par
\subsection*{Ultraviolet radiation of the wind}
We approximate the wind as a spherical gray body that emits photons of $E_\mathrm{\gamma,\,wind} = 100\:{\rm eV}$, and $L_\mathrm{\gamma,\, wind} = 10^{40}\:{\rm erg\,s^{-1}}$. Therefore, the rate of ionizing photons emitted by the wind is given by:
\begin{equation}
\dot{N}_\mathrm{UV,\, wind} = \frac{L_\mathrm{UV,\, wind}}{E_\mathrm{UV,\,wind}} \approx 10^{49}\;\mathrm{ph\,s^{-1}}.
\end{equation}
The size of the ionized zone by the wind can be obtained by the radius of Str\"{o}mgren $r_{\mathrm{st}}$ \citep[see e.g.][]{stromgren1939}, which is defined by equating the rate of ionizing photons to the recombination rate of electrons and protons $\dot{N}_{\mathrm{r}}$:
\begin{equation}
\dot{N}_{\mathrm{r}} = \frac{4}{3}\pi r_{\mathrm{st}}^{3} n_{\mathrm{med}}^{2}\beta_{2}(T),
\end{equation}
where $\beta_{2}$ is the recombination coefficient to just the excited states (a recombination directly to the ground state emits a Lyman continuum photon which is absorbed in ionizing another nearby H atom), $\beta_{2} \approx 2\times10^{-13}\,\mathrm{cm^{3}\,s^{-1}}$ for $T = 10^{4}\,\mathrm{K}$ \textbf{\citep[see e.g.][]{yu2005}.}
Characteristic values for the density of primordial star-forming clouds are $n_{\mathrm{med}} = 5\times 10^{3}\,{\rm cm^{-3}}$ \citep{bromm2002}. We obtain:
\begin{equation}
 r_{\mathrm{st,\,wind}} \sim 1\,\mathrm{kpc}
\end{equation}
Therefore, we conclude that the impact of the wind on reionization is similar to that of the first stars.\par
\subsection*{Neutron decay from inner jet}
Several channels of photohadronic interactions produce relativistic neutrons. Unlike charged particles, neutrons are not magnetically bounded by the jet and can escape. These particles can generate the injection of high-energy protons and low energy electrons in the environment. We now consider the role on the reionization of neutrons produced in the inner jet.\par
\begin{figure*}[h!]
   \centering
   \includegraphics[scale=1.1]{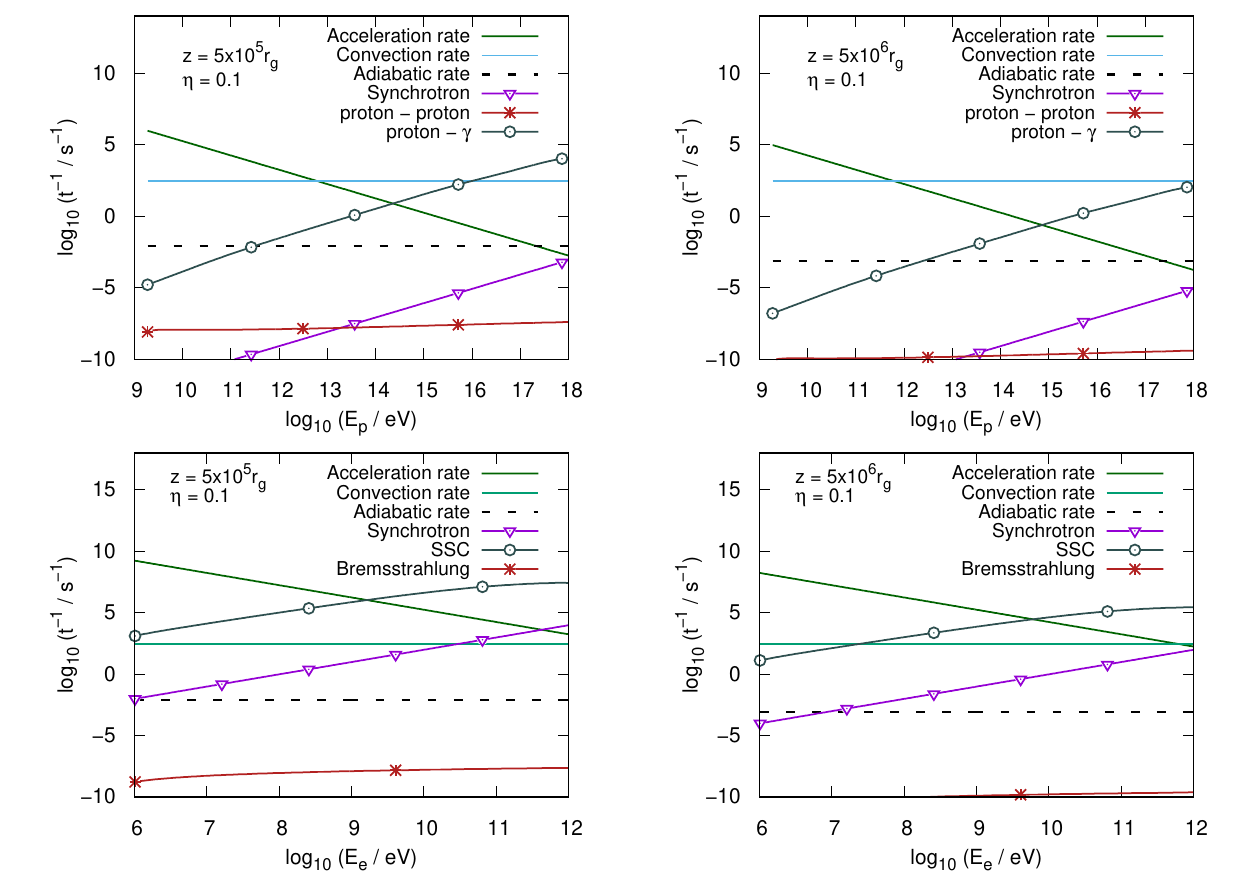}
      \caption{\small Cooling and acceleration times for populations of primary particles in the inner jet. We consider two cases for the location of the acceleration region. It is observed that while the protons reach a lower maximum energy when the acceleration region is located at a higher distance from the binary system, the electrons cool less efficiently. This is because of the difference between the dominant cooling processes for both particle populations.}
         \label{fig:TimesJet}
   \end{figure*}

Figure \ref{fig:TimesJet} shows the cooling, acceleration, and escape rates for relativistic electrons and protons. We consider an acceleration region located where the jet emerges from the wind photosphere, $z_{\rm acc} = 5\times 10^{5}\,r_{\rm g}$ (see the first box in Fig \ref{fig:TimesJet}). We see that the maximum energy of protons is $E_{\rm p}\sim 10\,{\rm TeV}$. In photohadronic interactions, neutrons are left with about 80\% of the proton energy and carry therefore cosmic ray energy efficiently \citep[see][]{rachen1998}.\par
The mean life-time of free neutrons in their rest-frame is $\tau_{\rm n} \approx 886\,{\rm s}$, and then they decay into protons, electrons, and antineutrinos. In the observer rest-frame, the decay rate is $t_{\rm dec} = \tau_{\rm n} E_{\rm n}/m_{\rm n}c^{2}$. We compare the decay time with the cooling time by neutron-proton radiative losses. For neutrons with $E_{\rm n}=10\,{\rm TeV}$ in a medium of $n = 5\times 10^{3} \,{\rm cm^-3}$, the characteristic cooling time is $t_{\rm np} \sim 10^{11}\,{\rm s}$. On the other hand, for the most energetic neutrons, we obtain $t_{\rm dec} \sim 10^{7}\,{\rm s}$, so we conclude that neutrons decay without interacting. 
If the acceleration region is located further away from the binary system, the protons reach a lower energy, and the neutrons decay faster. For $z_{\rm acc} = 5\times 10^{6}\,r_{\rm g}$, we obtain $t_{\rm dec} \sim 10^{6}\,{\rm s}$.
\par
Neutrons travel a distance of $d \approx 10^{14}\,{\rm cm}$, and then protons with $E \sim 10\,{\rm TeV}$ are injected  within of the dense medium in which the binary system was formed. These particles are cooled mainly by inelastic collisions with thermal protons. To determine if relativistic protons can escape the primordial halo, we calculate the mean free path between hadronic collisions. For $E = 10\,{\rm TeV}$, we obtain $\lambda = 1/\sigma_{\rm pp}n \sim 1\,{\rm kpc}$. Therefore, the escape of high-energy neutrons from the inner jet is an efficient mechanism of injection of non-relativistic electrons that collisionally ionize the local environment, and high-energy protons that initiate cascades at long distances from the binary system.\par
\subsection*{Relativistic electrons and protons from terminal jet}
High-energy electrons that escape from the terminal jets can ionize and heat the IGM, as discussed in \cite{douna2018}. Electrons with energy $E \lesssim 1\, {\rm GeV}$ escape from the terminal jet before being cooled inside the jet by synchrotron radiation and inverse Compton scattering. These electrons interact with the CMB, then decrease their energy and produce X-rays and gamma-rays. Gamma-rays are annihilated with the CMB, and the created pairs emit gamma-rays again. An electromagnetic cascade develops in the medium. The result of this cascade will be to degrade the energy of the original photons and multiply the number of leptons. The emerging spectrum will depend on the original spectrum of injection and the characteristics of the medium \citep[see][]{romero2011}. Electrons with kinetic energy of $E\sim 1\,{\rm keV}$ efficiently ionize neutral atoms by collisions. The detailed study of the production of low energy electrons by electromagnetic cascades requires the implementation of numerical simulations.\par
We have considered that in the terminal region of the jets the power in relativistic particles is a fraction $q=0.01$ of the kinetic power of the jet. In addition, the content of relativistic protons is the same as that of relativistic electrons ($a = L_{\mathrm{p}} / L_{\mathrm{e}} = 1$ in the terminal region). Protons are advected from the acceleration region long before radiative cooling. We can do an elemental analysis of the heating of the IGM by the protons in the terminal jets following \cite{jana2018} and \cite{sazonov2015} (heating of the IGM by SNR of Population III). The increase in the temperature of the IGM, $\Delta T_{\mathrm{IGM}}$, can be estimated by:
\begin{equation}
\frac{3}{2} \left( \frac{\rho_{\mathrm{IGM}}}{1.2\,m_{\mathrm{p}}} \right) \kappa_{\mathrm{B}}
\Delta T_{\mathrm{IGM}} = f_{\mathrm{heat}}
\eta_{\mathrm{p}}E_{\mathrm{jet}}n_{\mathrm{h}},
\end{equation}
where $f_{\mathrm{heat}} = 0.25$ is the fraction of proton kinetic energy deposited as heat through Coulomb interactions, $\eta_{\mathrm{p}} = 0.005$ is the fraction in relativistic protons of the kinetic power of the jet, $E_{\mathrm{jet}} = L_{\mathrm{jet}}\times t_{\mathrm{MQ}} \approx 3\times 10^{53}\,\mathrm{erg}$  is the total energy deposited in the IGM by the terminal jet, and $n_{\mathrm{h}} = 338\, (1 + z)^{3}\,\mathrm{Mpc}^{-3}$ is the comoving number density of the minihalo where the binary system was formed. Considering $z=10$ we obtain $\Delta T_{\mathrm{IGM}} = 13.9\,\mathrm{K}$. This effect is similar to that calculated for SNR of Population III at $z = 10$ in \cite{jana2018}, $\Delta T_{\mathrm{IGM,\,SNR}} \approx 15\,\mathrm{K}$ with $E_{\mathrm{SN}} = 10^{52}\,\mathrm{erg}$. These estimates are very rough and should be validated by numerical simulations.
 \par

\section{Discussion}
\label{sect:discussion}

From the results showed in the previous section we conclude that microquasars in early Universe could have had a non-negligible contribution on the reionization and heating of the intergalactic medium on long distances. Jets in microquasars of Population III can be significant $\gamma$-rays sources in the range MeV - GeV because of internal shocks in regions near to compact object. Emission from the jets in the range GeV - TeV should be produced in the terminal region where the internal and external absorption is negligible. Similarly, accelerated electrons in the terminal region of the jets can reach very-high energies, because the cooling mechanism by synchrotron radiation becomes inefficient.\par
We have considered a lepto-hadronic jet model in the near region to the compact object, where the power of relativistic particles is dominated by protons. It is easy to calculate lepton-dominated jet models. In such models it is expected a greater absorption caused by $\gamma \gamma$ annihilation with the synchrotron radiation field of primary electrons \citep{romero2008}.\par
For the calculation of the thermal spectral energy distribution of the accretion disk and wind we have considered a super-critical disk model \citep{fukue2004,akizuki2006}. The luminosity of the disk is regulated approximately to the value of the Eddington luminosity, because of the intense mass-loss by disk winds. The photons produced on the surface of the disk must interact with the ejected matter. Since the mass-loss is very intense (practically all the accreted material is expelled from the disk), it is expected that a large part of this radiation will be absorbed (e.g., by direct Compton effect), reducing its impact on the subsequent reionization of the intergalactic medium.\par
Another strong assumption that we have used in this paper is that the interaction between the jets and the wind is negligible. Therefore, jets emerge from the wind photosphere without decelerating by the pressure exerted on the jet head. The space of parameters should be explored such that scenarios of choked jets by the wind are considered or discarded. Another possible consequence of jet-wind interaction is the hadronic load in the jet. If the wind is fragmented into neutral clumps, they can penetrate the jet, and decelerate it. This is a possible way of hadronic load in the jets of SS433. These issues will be studied in a forthcoming paper using hydrodynamic simulations (Sotomayor Checa, Romero \& Bosch-Ramon, in prep.)\par
We have explored some mechanisms of reionization by microquasars. Electromagnetic cascades initiated by the cooling of relativistic protons seem to be a promising scenario. High-energy hadrons are injected by the decay of neutrons that escape from the inner jet. We have not taken into account the production of neutrons in collisions $pp$. This channel would be improved if neutral clumps penetrate the jet and collide with relativistic protons. A complete description of the jet-wind interaction is necessary. Relativistic electrons from the terminal jets also trigger electromagnetic cascades, but in regions of the IGM much farther from the binary system. It is expected that the highest contribution to the reionization will be provided by the jet terminal.\par
A detailed investigation of the amount reionization produced by microquasars of Population III requires to compute numerical simulations in order to calculate the production of photons and high-energy particles in electromagnetic cascades in the IGM. This last issue will also be studied exclusively in a forthcoming work.
\section{Conclusions}
\label{sect:conclusions}
We have developed a simple model for microquasars where the donor star is from Population III  placing special emphasis on the high-energy emission from these hypothetical sources. We have considered a proton-dominated microquasar that predicts significant gamma-ray emission and production of high-energy cosmic rays in the terminal regions of the jets. Within the jets, in the region of particle acceleration near the compact object, the effects of absorption by internal $\gamma\gamma$-annihilation are dramatic: radiation is completely suppressed for energies higher than MeV.\par
The basic assumptions of the model are that the mass-transfer is by spill of matter through the Langrange point $L_{1}$ and this occurs on a thermal time-scale. One of the main consequences of these hypotheses is that the mass accretion rate at the outer edge of the accretion disk is in the super-Eddington regime. This constitutes an important difference with respect to the observed galactic microquasars, except the enigmatic binary SS433.\par
Before the launch of jets in the first microquasars, the photons of highest energy are emitted in the accretion disk. However, the photons from the innermost region of the disk, which are those with the highest energy $E_{\gamma}\sim 10\,\mathrm{keV}$, are attenuated by the wind. Therefore, the binary system is a source of a lot of ultraviolet radiation.\par 
An elementary study allows us to infer that microquasars in the early Universe could have participated to the reionization of the IGM through several processes. The ultraviolet radiation of the star, disk, and wind ionizing the environment of the system, but having no effect on the IGM at long distances. Relativistic neutrons are created in the inner jet by photohadronic interactions. These particles escape from the jet and decay into very high-energy protons, non-relativistic electrons, and neutrinos. Electrons ionize the surrounding medium colisionally. Production of neutral pions by the cooling of relativistic protons is expected. Annihilation of gamma-rays produced by pion decay create pairs. Relativistic leptons again produce gamma-rays by inverse Compton scattering. An electromagnetic cascade is initiated and it injects ionizing photons in the IGM. In all cases the photon target is the CMB. Similarly, accelerated protons that escape the terminal jet drive electromagnetic cascades in the IGM. Reionization by the broadband radiation emitted in the jet has not been considered in this paper and require an additional treatment. The next step in this work is to analyze these mechanisms in detail using numerical simulations for the study of electromagnetic cascades. An additional prediction of our model is the production of high-energy neutrinos. These issues will be included in forthcoming communications.\par
\begin{acknowledgements}
We thank an anonymous referee for his insightful comments and suggestions that improved this paper. This work was support with grants PIP 0338 (CONICET). PSC is grateful with the National University of La Plata and the public education system in Argentina for the training received. GER acknowledges support by the Ministerio de Econom\'ia y Competitividad (MINECO) under grant AYA2016-76012-C3-1-P. 
\end{acknowledgements}
%

\begin{thebibliography}{aa}



\bibitem[Aharonian(2004)]{aharonian2004} Aharonian, F.~A.\ 2004, Very High Energy Cosmic Gamma Radiation: A Crucial Window on the Extreme Universe. ~Published by World Scientific Publishing Co.~Pte.~Ltd., .~ISBN \#9789812561732, 9789812561732 





\bibitem[Akizuki \& Fukue(2006)]{akizuki2006} Akizuki, C., \& Fukue, J.\ 2006, \pasj, 58, 469 



\bibitem[Atoyan \& Dermer(2003)]{atoyan2003} Atoyan, A.~M., \& Dermer, C.~D.\ 2003, \apj, 586, 79 

\bibitem[Bahena \& Hadrava(2012)]{bahena2012} Bahena, D., \& Hadrava, P.\ 2012, \apss, 337, 651 

\bibitem[Bahena \& Klapp(2010)]{bahena2010} Bahena, D., \& Klapp, J.\ 2010, \apss, 327, 219 




\bibitem[Begelman(1978)]{begelman1978} Begelman, M.~C.\ 1978, \mnras, 184, 53 

\bibitem[Begelman et al.(1990)]{begelman1990} Begelman, M.~C., Rudak, B., \& Sikora, M.\ 1990, \apj, 362, 38 

\bibitem[Beloborodov(1998)]{beloborodov1998} Beloborodov, A.~M.\ 1998, \mnras, 297, 739 

\bibitem[Berezinskii et al.(1990)]{berezinskii1990} Berezinskii, V.~S., Bulanov, S.~V., Dogiel, V.~A., \& Ptuskin, V.~S.\ 1990, Amsterdam: North-Holland, 1990, edited by Ginzburg, V.L.,  


\bibitem[Blumenthal \& Gould(1970)]{blumenthal1970} Blumenthal, G.~R., \& Gould, R.~J.\ 1970, Reviews of Modern Physics, 42, 237 

\bibitem[Bordas et al.(2009)]{bordas2009} Bordas, P., Bosch-Ramon, V., Paredes, J.~M., \& Perucho, M.\ 2009, \aap, 497, 325 

\bibitem[Bosch-Ramon(2018)]{bosch-ramon2018} Bosch-Ramon, V.\ 2018, arXiv:1808.08911 

\bibitem[Bosch-Ramon et al.(2006)]{bosch-ramon2006} Bosch-Ramon, V., Romero, G.~E., \& Paredes, J.~M.\ 2006, \aap, 447, 263 


\bibitem[Bromm et al.(2002)]{bromm2002} Bromm, V., Coppi, P.~S., \& Larson, R.~B.\ 2002, \apj, 564, 23 

\bibitem[Bromm(2013)]{bromm2013} Bromm, V.\ 2013, Reports on Progress in Physics, 76, 112901 

\bibitem[Calvani \& Nobili(1981)]{calvani1981} Calvani, M., \& Nobili, L.\ 1981, \apss, 79, 387 






\bibitem[Douna et al.(2018)]{douna2018} Douna, V.~M., Pellizza, L.~J., Laurent, P., \& Mirabel, I.~F.\ 2018, \mnras, 474, 3488 

\bibitem[Drury(1983)]{drury1983} Drury, L.~O.\ 1983, Reports on Progress in Physics, 46, 973 

\bibitem[Dubus(2006)]{dubus2006} Dubus, G.\ 2006, \aap, 451, 9 

\bibitem[Eggleton(1983)]{eggleton1983} Eggleton, P.~P.\ 1983, \apj, 268, 368 

\bibitem[Ellis et al.(2012)]{ellis2012} Ellis, G.~F.~R., Maartens, R., \& MacCallum, M.~A.~H.\ 2012, Relativistic Cosmology, by George F.~R.~Ellis , Roy Maartens , Malcolm A.~H.~MacCallum, Cambridge, UK: Cambridge University Press, 2012,

\bibitem[Fabrika(2004)]{fabrika2004} Fabrika, S.~N.\ 2004, Astrophysics and Space Physics Reviews, 12, 1

\bibitem[Fabrika et al.(2007)]{fabrika2007} Fabrika, S.~N., Abolmasov, P.~K., \& Karpov, S.\ 2007, Black Holes from Stars to Galaxies -- Across the Range of Masses, 238, 225 




\bibitem[Fukue(2000)]{fukue2000} Fukue, J.\ 2000, \pasj, 52, 829  

\bibitem[Fukue(2004)]{fukue2004} Fukue, J.\ 2004, \pasj, 56, 569 

\bibitem[Fukue \& Iino(2010)]{fukue2010} Fukue, J., \& Iino, E.\ 2010, \pasj, 62, 1399 

\bibitem[Fukue \& Sumitomo(2009)]{fukue2009} Fukue, J., \& Sumitomo, N.\ 2009, \pasj, 61, 615 



\bibitem[Heger et al.(2003)]{heger2003} Heger, A., Fryer, C.~L., Woosley, S.~E., Langer, N., \& Hartmann, D.~H.\ 2003, \apj, 591, 288 

\bibitem[Inayoshi et al.(2017)]{inayoshi2017} Inayoshi, K., Hirai, R., Kinugawa, T., \& Hotokezaka, K.\ 2017, \mnras, 468, 5020 

\bibitem[Jana et al.(2018)]{jana2018} Jana, R., Nath, B.~B., \& Biermann, P.~L.\ 2018, \mnras,  

\bibitem[Jaroszynski et al.(1980)]{jaroszynski1980} Jaroszynski, M., Abramowicz, M.~A., \& Paczynski, B.\ 1980, \actaa, 30, 1 


\bibitem[Khangulyan et al.(2007)]{khangulyan2007} Khangulyan, D., Hnatic, S., Aharonian, F., \& Bogovalov, S.\ 2007, \mnras, 380, 320 

\bibitem[Kelner et al.(2006)]{kelner2006} Kelner, S.~R., Aharonian, F.~A., \& Bugayov, V.~V.\ 2006, \prd, 74, 034018 


\bibitem[Kitabatake et al.(2002)]{kitabatake2002} Kitabatake, E., Fukue, J., \& Matsumoto, K.\ 2002, \pasj, 54, 235 

\bibitem[Krti{\v c}ka \& Kub{\'a}t(2006)]{krtica2006} Krti{\v c}ka, J., \& Kub{\'a}t, J.\ 2006, Stellar Evolution at Low Metallicity: Mass Loss, Explosions, Cosmology, 353, 133 



\bibitem[Liska et al.(2018)]{liska2018} Liska, M.~T.~P., Tchekhovskoy, A., \& Quataert, E.\ 2018, arXiv:1809.04608 

\bibitem[Loeb(2010)]{loeb2010} Loeb, A.\ 2010, How Did the First Stars and Galaxies Form? By Abraham Loeb.~Princeton University Press, 2010.~ISBN: 978-1-4008-3406-8,  

\bibitem[Madau et al.(2004)]{madau2004} Madau, P., Rees, M.~J., Volonteri, M., Haardt, F., \& Oh, S.~P.\ 2004, \apj, 604, 484 




\bibitem[Meier(2005)]{meier2005} Meier, D.~L.\ 2005, \apss, 300, 55 

\bibitem[Milosavljevi{\'c} et al.(2009)]{milosavljevic2009} Milosavljevi{\'c}, M., Couch, S.~M., \& Bromm, V.\ 2009, \apjl, 696, L146 


\bibitem[Mirabel et al.(2011)]{mirabel2011} Mirabel, I.~F., Dijkstra, M., Laurent, P., Loeb, A., \& Pritchard, J.~R.\ 2011, \aap, 528, A149 


\bibitem[Mirabel \& Rodr{\'{\i}}guez(1994)]{mirabel1994} Mirabel, I.~F., \& Rodr{\'{\i}}guez, L.~F.\ 1994, \nat, 371, 46 

\bibitem[Mirabel \& Rodr{\'{\i}}guez(1999)]{mirabel1999} Mirabel, I.~F., \& Rodr{\'{\i}}guez, L.~F.\ 1999, \araa, 37, 409 


\bibitem[Narayan et al.(2003)]{narayan2003} Narayan, R., Igumenshchev, I.~V., \& Abramowicz, M.~A.\ 2003, \pasj, 55, L69 

\bibitem[Narayan \& Yi(1994)]{narayan1994} Narayan, R., \& Yi, I.\ 1994, \apjl, 428, L13 


\bibitem[Ohsuga et al.(2003)]{ohsuga2003} Ohsuga, K., Mineshige, S., \& Watarai, K.-y.\ 2003, \apj, 596, 429 

\bibitem[Ohsuga et al.(2005)]{ohsuga2005} Ohsuga, K., Mori, M., Nakamoto, T., \& Mineshige, S.\ 2005, \apj, 628, 368 

\bibitem[Paczy{\'n}ski(1971)]{paczynski1971} Paczy{\'n}ski, B.\ 1971, \araa, 9, 183 

\bibitem[Paczy{\'n}sky \& Wiita(1980)]{paczynski1980} Paczy{\'n}sky, B., \& Wiita, P.~J.\ 1980, \aap, 88, 23 




\bibitem[Piran(1982)]{piran1982} Piran, T.\ 1982, \apjl, 257, L23 

\bibitem[Qin et al.(2017)]{qin2017} Qin, Y., Mutch, S.~J., Poole, G.~B., et al.\ 2017, \mnras, 472, 2009 

\bibitem[Rachen \& M{\'e}sz{\'a}ros(1998)]{rachen1998} Rachen, J.~P., \& M{\'e}sz{\'a}ros, P.\ 1998, Gamma-Ray Bursts, 4th Hunstville Symposium, 428, 776 

\bibitem[Reynoso, Christiansen, \& Romero(2008)]{reynoso2008a} Reynoso, M.~M., Christiansen, H.~R., \& Romero, G.~E.\ 2008, Astroparticle Physics, 28, 565 


\bibitem[Reynoso \& Romero(2009)]{reynoso2009} Reynoso, M.~M., \& Romero, G.~E.\ 2009, \aap, 493, 1 

\bibitem[Reynoso, Romero, \& Christiansen (2008)]{reynoso2008b} Reynoso, M.~M., Romero, G.~E., \& Christiansen, H.~R.\ 2008, \mnras, 387, 1745 


\bibitem[Romero, del Valle, \& Orellana(2010)]{romero2010b} Romero, G.~E., del Valle, M.~V., \& Orellana, M.\ 2010, \aap, 518, A12 

\bibitem[Romero et al.(2007)]{romero2007} Romero, G.~E., Okazaki, A.~T., Orellana, M., \& Owocki, S.~P.\ 2007, \aap, 474, 15 

\bibitem[Romero \& Paredes(2011)]{romero2011} Romero, G.~E., \& Paredes, J.~M.\ 2011, Introducci\'on a la Astrof\'isica Relativista, Publications i Edicions de la Universitat de Barcelona,

\bibitem[Romero \& Sotomayor Checa(2018)]{romero2018} Romero, G.~E., \& Sotomayor Checa, P.\ 2018, International Journal of Modern Physics D, 27, 1844019 

\bibitem[Romero, Vieyro \& Vila(2010)]{romero2010a} Romero, G.~E., Vieyro, F.~L., \& Vila, G.~S.\ 2010, \aap, 519, A109 

\bibitem[Romero \& Vila(2008)]{romero2008} Romero, G.~E., \& Vila, G.~S.\ 2008, \aap, 485, 623 

\bibitem[Romero \& Vila(2014)]{romero2014} Romero, G.~E., \& Vila, G.~S.\ 2014, Introduction to Black hole Astrophysics, Lecture Notes in Physics, 876, Berlin Springer Verlag


\bibitem[S{\c a}dowski \& Narayan(2015)]{sadowski2015} S{\c a}dowski, A., \& Narayan, R.\ 2015, \mnras, 453, 3213 


\bibitem[Sazonov \& Sunyaev(2015)]{sazonov2015} Sazonov, S., \& Sunyaev, R.\ 2015, \mnras, 454, 3464 

\bibitem[Shakura \& Sunyaev(1973)]{shakura1973} Shakura, N.~I., \& Sunyaev, R.~A.\ 1973, \aap, 24, 337 


\bibitem[Sotomayor Checa \& Romero(2019)]{sotomayor2019a} Sotomayor Checa, P., \& Romero, G.~E.\ 2019, arXiv:1902.04670

\bibitem[Spada et al.(2001)]{spada2001} Spada, M., Ghisellini, G., Lazzati, D., \& Celotti, A.\ 2001, \mnras, 325, 1559 


\bibitem[Stecker(1968)]{stecker1968} Stecker, F.~W.\ 1968, Physical Review Letters, 21, 1016 

\bibitem[Str{\"o}mgren(1939)]{stromgren1939} Str{\"o}mgren, B.\ 1939, \apj, 89, 526 



\bibitem[Tueros et al.(2014)]{tueros2014} Tueros, M., del Valle, M.~V., \& Romero, G.~E.\ 2014, \aap, 570, L3 

\bibitem[Vieyro \& Romero(2012)]{vieyro2012} Vieyro, F.~L., \& Romero, G.~E.\ 2012, \aap, 542, A7 

\bibitem[Vila(2012)]{vilatesis} Vila, G.~S.\ 2012, Ph.D.~Thesis  




\bibitem[Watarai \& Fukue(1999)]{watarai1999} Watarai, K.-y., \& Fukue, J.\ 1999, \pasj, 51, 725 

\bibitem[Wiita(1982)]{wiita1982} Wiita, P.~J.\ 1982, \apj, 256, 666 


\bibitem[Yu(2005)]{yu2005} Yu, Q.\ 2005, \apj, 623, 683 

\bibitem[Zhou et al.(2018)]{zhou2018} Zhou, Y., Feng, H., Ho, L.~C., \& Yao, Y.\ 2018, arXiv:1812.02923 

\end{thebibliography}
%

\end{document}